\documentclass[twocolumn,eqsecnum,floatfix,aps,prc]{revtex4}
\usepackage{epsfig}

\def\lap{\mathrel{\mathpalette\fun <}}

\def\fun#1#2{\lower3.6pt\vbox{\baselineskip0pt\lineskip.9pt
  \ialign{$\mathsurround=0pt#1\hfil##\hfil$\crcr#2\crcr\sim\crcr}}}
\begin{document}

%\parindent=10pt
%%%%%%%%%%%%%%%%%%%%%%%%%%%%%%%%%%%%%%%%%%%%%%%%%%%%%%%%%%%%%%%%

\title { Variational calculation of $^4$H\lowercase{e} tetramer ground
and excited states \\ 
%\vskip 0.2cm
using a realistic pair potential
}

\author{E.\ Hiyama}
\email{hiyama@riken.jp}
\affiliation{RIKEN Nishina Center, RIKEN, Wako 351-0198, Japan}

\author{M.\ Kamimura}
\email{mkamimura@riken.jp}
\affiliation{Department of Physics, Kyushu University,
Fukuoka 812-8581, Japan, \\
RIKEN Nishina Center, RIKEN, Wako 351-0198, Japan}

\date{\today}

%%%%%%%%%%%%%%%%%%%%%%%%%%%%
\begin{abstract}
We calculated the $^4$He  trimer and tetramer ground and excited states
with the LM2M2 potential 
using our Gaussian expansion method (GEM)
for {\it ab initio} variational calculations of few-body systems.
The method has extensively been used for %  in the studies of
a variety of three-, four- and five-body
systems in nuclear physics and exotic atomic/molecular physics.
The trimer (tetramer) wave function is expanded in terms of
symmetric three-(four-)body Gaussian basis functions, 
ranging from very compact to very diffuse,
without assuming any  pair correlation function.
Calculated results of the trimer ground and excited states
are in excellent agreement with the literature.
Binding energies of the tetramer
ground and excited states are obtained 
to be 558.98 mK and 127.33 mK (0.93 mK below the 
trimer ground state), respectively. 
We found that precisely the same shape of the short-range 
correlation ($ r_{ij} \lap 4 $\AA) in  the dimer appear in 
the ground and excited states of trimer and tetramer. 
The overlap function  between
the trimer excited state and the dimer  
and that between the tetramer excited state and the trimer
ground state are almost proportional to the dimer wave function
in the asymptotic region (up to $\sim 1000$ \AA).
Also the pair correlation functions
of trimer and tetramer excited states 
are almost proportional to
the squared dimer wave function. 
We then come to propose a  model 
which predicts the  binding energy of the first excited state
of $^4$He$_N$  ($N \ge 3$) measured from the $^4$He$_{N-1}$ ground state
to be nearly $\frac{N}{2(N-1)} B_2$ using the
dimer binding energy $B_2$.
\end{abstract}

   
\maketitle

\section{INTRODUCTION}

In early 1970's,  Efimov pointed out 
a possibility of having an infinite 
number of three-body bound states even
when none exists in the separate two-body 
subsystems~\cite{Efimov70,Efimov73,Efimov11}. 
This occurs when the two-body
scattering length is much larger than  the range of the
two-body interaction.
As a candidate of such three-body states,
Efimov discussed about the famous Hoyle state~\cite{Hoyle} 
(the second $0^+$ state at 7.65MeV
in the  $^{12}$C nucleus) taking a model of three
$\alpha$ particles (clusters of three $^4$He nuclei) 
as well as about the three-nucleon bound state ($^3$H nuclei).
In nuclear systems, the Borromean states, 
weakly bound three-body 
states though having no bound two-body subsystems,
are familiar but not classified as  Efimov states.
%% so far no evidence of nuclear Efimov states have been found.

%\vskip 0.1cm
In atomic systems, 
triatomic $^4$He (trimer) 
have been expected to have bound states of Efimov type since 
the realistic $^4$He-$^4$He 
interactions~\cite{HFDHE2,LM2M2,TTY,SAPT2,HFDB3FCI1}
give a large $^4$He-$^4$He scattering length ($\simeq 115$\AA), 
much greater than the potential range ($\sim 10$\AA),
and a very small $^4$He dimer binding energy ($\simeq 1.3$ mK).
(Experimentally, Ref.~\cite{Grisenti2000} evaluated a scattering length
of $104^{+8}_{-18}$ \AA$\,$ and 
a binding energy of $1.1^{+0.3}_{-0.2}$ mK).

%\vskip 0.1cm
As is mentioned in recent reviews about 
the $^4$He trimer~\cite{review-2009,review-Efimov}
(further references therein),
i) a lot of three-body calculations using the realistic pair potentials
have shown that the
$^4$He trimer possesses two bound states with binding energies of
nearly $126.4$ mK and $2.3$ mK,
\mbox{ii) it is already} rather well established that, 
if the  $^4$He trimer excited state exist, it should be Efimov nature,
and iii) it is suggested that the $^4$He trimer ground state 
may be considered as an Efimov state 
since the ground- and excited-state binding energies
move along the same universal scaling curve under any small deformation 
of the two-body potential (for details, see, e.g., 
Sec.III of Ref.~\cite{Braaten03}).
Experimentally, the $^4$He trimer ground state has been
observed  in Ref.~\cite{dimer-exp2005} 
to have the $^4$He-$^4$He bond length 
of $11^{+4}_{-5}$ \AA$\,$ in agreement with theoretical predictions,
whereas a reliable experimental evidence for the $^4$He trimer
excited state is still missing.

%\vskip 0.1cm
Only very recently, experimental evidences 
of Efimov trimer states have been reported
in the work using the ultracold  gases
of cesium atoms~\cite{cesium-Kramer,cesium-Knoop}, 
potasium atoms~\cite{Potasium-Zaccanti}, 
lithium-7 atoms~\cite{Lithium-Gross,Lithium-Pollack}, 
and \mbox{lithium-6} atoms~\cite{Li6-Lompe,Li6-Nakajima1,
Li6-Nakajima2,Li6-Ottensen,Li6-Williams},
in which the two-body interaction between 
those alkali atoms was manipulated 
so as to tune the scattering length to values 
significantly greater than the potential range.
These experiments have been accessing the study of
a wide variety of interesting physical systems 
in the atomic and nuclear fields.
Recently, the study extends to 
the Efimov physics and universality 
of four-atomic systems (tetramers).

%\vskip 0.1cm
Though the interactions between $^4$He atoms can not
be manipulated,
the study of $^4$He trimer using the realistic pair potentials
has been providing fundamental information to the Efimov physics.
Now it is  one of the challenging subjects
to precisely investigate the structure of
$^4$He tetramer using the realistic $^4$He-$^4$He
potential.

%\vskip 0.1cm
So far there
exist in the literature a large number of
$^4$He trimer calculations~\cite{Carbonell,Kievsky01,Roudnev,
Lewerenz,Bressanini,Blume,Das,Pandha83,Gloeckle86,
Carbonell93,Nielsen98,Roudnev00,Motovilov01,Kologanova04,Suno08}
giving well converged results with the
realistic $^4$He-$^4$He interactions.
However, calculations of the tetramer 
remain limited~\cite{Carbonell,Lewerenz,Bressanini,Blume,Das};
in those papers, the  binding energy of the tetramer ground state  
agrees well with each other, while
that of the loosely bound excited state  differs  
significantly from one another.

%\vskip 0.1cm
Thus the main purpose of the present paper is
to perform accurate calculations
of the $^4$He tetramer ground and excited states
using a realistic $^4$He-$^4$He interaction, 
the LM2M2 potential~\cite{LM2M2}.
We employ the Gaussian expansion method (GEM)
for {\it ab initio} variational calculations of 
few-body systems~\cite{Kamimura88,Kameyama89,Kamimura90,Hiyama03}.
The method has been proposed and developed by 
the present authors and collaborators
and applied to various types of three-,
four- and five-body systems in nuclear physics and 
exotic atomic/molecular physics 
(cf. review papers \cite{Hiyama03,Hiyama09,Hiyama10}).

%\vskip 0.1cm
Advantage  of using  the GEM for the $^4$He tetramer calculation
in the presence of the strong short-range repulsive
potential is as follows:
Some 30000 symmetrized four-body Gaussian basis functions, 
ranging from very compact to very diffuse, 
are constructed on the full 18 sets of Jacobi coordinates  
without assuming any pair correlation function.
They forms a nearly complete set in the finite coordinate space
concerned, so that one can describe accurately both the
short-range structure and the long-range asymptotic behavior 
(up to $\sim 1000$ \AA) of the four-body wave function, which
makes it possible to find new facets of  $^4$He clusters.

%\vskip 0.1cm
We thus find that precisely 
the same shape of the short-range
correlation ($r_{ij} \lap 4$ \AA) in  dimer appears 
in the ground and excited states of trimer and tetramer.
This gives a foundation to
an {\it a priori} assumption that a two-particle 
correlation function (such as the Jastrow's) 
so as to simulate the short-range part of
the dimer wave function is incorporated in the trimer and tetramer
wave functions from the beginning.

%\vskip 0.1cm
By illustrating the asymptotic behavior 
of the $^4$He trimer and tetramer, 
we discuss about an interesting relation between
their excited-state wave functions and the dimer wave function.
We then come to propose 
a 'dimerlike-pair' model that predicts the binding energy 
of the first excited state of 
the $N$-cluster system, $^4$He$_N$, measured from the
ground state of $^4$He$_{N-1}$ to be approximately 
$\frac{N}{2(N-1)} B_2$
using the dimer binding energy $B_2$.

%\vskip 0.1cm
We explicitly write the asymptotic form of 
the total wave function of  $^4$He trimer (tetramer).
The asymptotic normalization 
coefficient (ANC)~\cite{Friar,Kameyama89,Kievsky01,ANC1,ANC2},
namely the amplitude of  tail function
of the \mbox{dimer-atom} (trimer-atom) relative motion
in the present case,
is a quantity to reflect the internal structure
of trimer (tetramer).
Therefore, attention to the ANC might be useful 
when one intends to reproduce the non-universal variation of the 
$^4$He trimer (tetramer) states 
by means of parametrizing \mbox{effective} models
beyond Efimov's universal theory.

The paper is organized as follows:
In Sec. II, we apply the GEM to the three-body calculation of
the $^4$He trimer ground and excited states showing that
the calculated results agree excellently with the literature.
In Sec. III, the four-body calculation of the $^4$He tetramer
ground and excited states is presented.
Summary is given in Sec. IV.

\section{$^4$H\lowercase{e} trimer}

The $^4$He trimer bound states have extensively been
studied in many theoretical work using realistic potentials.
Monte-Carlo, hyperspherical, variational 
and Faddeev techniques were used 
to calculate accurately the binding energies of 
the ground and excited 
states~\cite{Carbonell,Kievsky01,Roudnev,Lewerenz,Bressanini,Blume,Das,
Pandha83,Gloeckle86,Carbonell93,Nielsen98,Roudnev00,Suno08,
Motovilov01,Kologanova04}
 (see also recent reviews~\cite{review-2009,review-Efimov}).
Nevertheless,
in this section, we explain our Gaussian expansion method (GEM)
and present the calculated result 
for the $^4$He trimer in order to demonstrate
high accuracy of our calculation 
before we report our investigation of the $^4$He tetramer
in the next section.

\subsection{Three-body wave function}

\begin{figure}[h]
\begin{center}
\epsfig{file=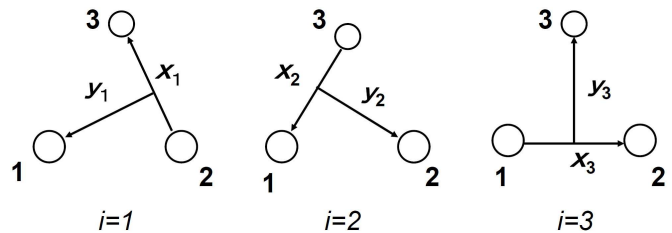,width=250pt}
\end{center}
\caption{
Three sets of the Jacobi coordinates for  $^4$He trimer.
}
\label{fig:3bodyjacobi}
\end{figure}

We take all the three sets of 
Jacobi coordinates (Fig.~\ref{fig:3bodyjacobi}),
${\bf x}_{1}={\bf r}_2 -{\bf r}_3$ and 
${\bf y}_{1}={\bf r}_1 -\frac{1}{2}({\bf r}_2 + {\bf r}_3)$ 
and cyclically for $({\bf x}_2, {\bf y}_2)$ and $({\bf x}_3, {\bf y}_3)$,
${\bf r}_i$ being the position vector of $i$th particle. 
Hamiltonian of the system is expressed as
\begin{equation}
H= -\frac{\hbar^2}{2\mu_{x}} \nabla^{2}_{x}
   -\frac{\hbar^2}{2\mu_{y}} \nabla^{2}_{y}
 + \sum_{1=i<j}^3 V(r_{ij}),
\label{eq:Hamil3}
\end{equation}
where 
$\mu_x=\frac{1}{2} m$ and $\mu_y=\frac{2}{3}m$, $m$ being  mass
of a $^4$He atom.
$V(r_{ij})$ is the two-body $^4$He-$^4$He potential
as a function of the pair distance 
${\bf r}_{ij}={\bf r}_j- {\bf r}_i$.

%%%%%%%%%%%%%%%%%%  Fig. 1 (Jacobi) %%%%%%%%%%%
%%%%%%%%%%%%%%%%%%%%%%%%%%%%%%%%%%%%%%%%%%%%%%%%%
We calculate the three-body bound-state wave function, $\Psi_3$,
which satisfies the Schr\"{o}dinger equation
\begin{equation}
( H - E ) \Psi_3 =0 .
\end{equation}
Since we consider the $^4$He atom as a spinless boson,
we expand the wave function of three identical spinless bosons
in terms of $L^2$-integrable, fully symmetric three-body basis 
functions:
\begin{eqnarray}
&& \quad \Psi_3 = \sum_{\alpha=1}^{\alpha_{\rm max}} 
A_\alpha \Phi_\alpha^{\rm (sym)}, \\
 \Phi_\alpha^{\rm (sym)} &=&
    \Phi_\alpha( {\bf x}_1, {\bf y}_1) 
 +  \Phi_\alpha( {\bf x}_2, {\bf y}_2) 
 +  \Phi_\alpha( {\bf x}_3, {\bf y}_3). \qquad 
\end{eqnarray}
It is of importance that those basis functions 
\{$\Phi_\alpha^{\rm (sym)}; \alpha=1, ..., \alpha_{\rm max}$\}, 
which are nonorthogonal to each other, are constructed 
on the full three sets of
Jacobi coordinates; this makes the function space 
of $\{ \Phi_\alpha^{\rm (sym)}$\} quite wide.

The eigenenergies $E$ and amplitudes   
$A_\alpha$ of the ground and excited states are determined
by the Rayleigh-Ritz variational principle:
\begin{equation}
\langle \: \Phi_{\alpha}^{\rm (sym)} \:| 
\:H - E \: |\: \Psi_3 \: \rangle =0 ,
\end{equation}
where $\alpha= 1, ...,  \alpha_{\rm max}$. Eqs.(2.5) results 
in a generalized eigenvalue problem:
\begin{equation}
 \sum_{\alpha'=1}^{\alpha_{\rm max}} \big[ {\cal H}_{\alpha,\alpha'} -
 E\, {\cal N}_{\alpha, \alpha'} \big] \,A_{\alpha'} = 0 .
\qquad 
\end{equation}
The matrix elements are written as
\begin{eqnarray}
 {\cal H}_{\alpha,  \alpha'} &=&  
\langle \: \Phi_{\alpha}^{\rm (sym)}  \:| \:H\: |\: 
\Phi_{\alpha'}^{\rm (sym)} \: \rangle ,\\
 {\cal N}_{\alpha,  \alpha'} &=&  
\langle \: \Phi_{\alpha}^{\rm (sym)}  \:| \:\: 1 \:\: |\: 
\Phi_{\alpha'}^{\rm (sym)} \: \rangle .
\end{eqnarray}
The lowest-lying two $S$-wave eigenstates,
$\Psi_3^{(v)} (v=0, 1)$, will be identified as
the trimer ground $(v=0)$ and excited ($v=1$)  states.

We express each basis function
$\Phi_\alpha ( {\bf x}_i, {\bf y}_i)$  
as a product of a function of ${\bf x}_i$ and 
that of ${\bf y}_i$:
%%%%%%%%%%%%%%%%%%%%%%%%%
\begin{equation}
 \Phi_\alpha( {\bf x}_i, {\bf y}_i)=
 \phi_{n_x l_x}(x_i)
        \psi_{n_y l_y}(y_i) 
\Big[ Y_{l_x}({\widehat {\bf x}_i}) %\otimes
Y_{l_y}({\widehat {\bf y}_i})
  \Big]_{J M},
\end{equation}
%%%%%%%%%%%%%%%%%%%%%%%%%%%%%%%%
where $\alpha$ specifies a set of
quantum numbers
\begin{eqnarray*}  %\begin{equation}
  \alpha = \{n_x l_x, n_y l_y ,JM \} % \nonumber
\end{eqnarray*}  %\end{equation}
commonly for the components  $i= 1, 2, 3$.
$J$ is the total angular momentum  and
$M$ is its $z$-component. 
In this paper, we consider the trimer bound states with $J=0$.
Then, the totally symmetric three-body wave function
requires $l_x=l_y={\rm even}$.

%\vskip 0.1cm
One of the most important issues of 
the present variational calculation
is what type of radial shape we use for 
$\phi_{n_x l_x}(x_i)$ and $\psi_{n_y l_y}(y_i)$.
The basis functions should be capable of precisely describing 
the strong short-range correlation (without assuming any correlation 
function {\it a priori}) and the long-range asymptotic behavior
of very loosely bound states. 

%\vskip 0.1cm
The GEM recommends two types of 
functions which are  tractable in few-body calculations
and work acculately.
One is the Gaussian function and the other, more powerful one,
is the complex-range Gaussian function~\cite{Hiyama03}.
In the next subsection, we introduce the former that was successfully
used in our previous study (Sec. 3.1 of Ref.~\cite{Hiyama03})
of the $^4$He trimer ground and
excited state with the use of the HFDHE2 potential~\cite{HFDHE2}.
The latter function is introduced in Sec.II.C.

%%%%%%%%%%%%%%%%%%%%%%%%%%%%%%%%%%%%%%%%%%%%
\subsection{Gaussian basis functions}
%%%%%%%%%%%%%%%%%%%%%%%%%%%%%%%%%%%%%%%%%%%%

The radial function $\phi_{n_x l_x}(x)$ in (2.9)  
is taken to be a Gaussian multiplied by $x^{l_{n_x}}$ 
(similarly for $\psi_{n_y l_y}(y)$ ):
\begin{eqnarray}
&&\phi_{n_x l_x}(x) =x^{l_x}\:e^{-(x/x_{n_x})^2},\: \;   \\
&&\psi_{n_y l_y}(y)=y^{l_y}\:e^{-(y/y_{n_y})^2}, \: \;   
\end{eqnarray}
where   normalization constants  are
omitted for simplicity. 
 
Setting of the ranges by stochastic or random choice
does not seem suitable for describing the strong short-range
correlation and the long-range asymptotic behavior of the wave function.
Any intended choice of the ranges is  necessary. The
GEM recommends to set them in a {\it geometric} progression:
\begin{eqnarray}
&& x_{n_x}= x_1\, a_x^{n_x-1}
\quad \:(n_x=1, ..., n_x^{\rm max})\;,  \\
&& y_{n_y}=y_1\, a_y^{n_y-1}
\quad \:(n_y=1, ..., n_y^{\rm max}),\;  
\end{eqnarray}
with common ratios $a_x>1$ and $a_y>1$.
This greatly reduces 
the nonlinear parameters to be optimized.
We designate a set of
the geometric sequence  
by $\{ n_x^{\rm max},\, x_1, \,
x_{n_x^{\rm max}} \}$  instead of   
$\{ n_x^{\rm max},\, x_1, \,a_x  \}$ 
and similarly for  $\{ n_y^{\rm max},\, y_1, \,
y_{n_y^{\rm max}} \}$ , which is more convenient
to consider the spatial distribution of the basis set.
Optimization of the nonlinear range parameters is in principle
by trial and error procedure but much of experiences and
systematics have been accumulated in many studies using the GEM.

%\vskip 0.1cm
The basis functions \{$\phi_{n l}$\}
have the following properties:
i) They range from very compact to very diffuse, more densely 
in the inner region than in the outer one.
While the basis functions with small ranges are responsible for describing 
the short-range structure of the system, the basis 
with longest-range parameters are for the asymptotic behavior. 
ii) They, being multiplied by normalization
constants for $\langle \phi_{n l} \,|\, \phi_{n l} \rangle =1$,
have a relation
\begin{equation}
 \langle \phi_{n \, l} \,|\, \phi_{n+k \, l} \rangle =
\left( \frac{2a^k}{1+a^{2k}} \right)^{l+3/2} ,
\end{equation}
which tells that the overlap with the $k$-th neighbor
is {\it independent} of $n$, decreasing gradually with increasing $k$.

%\vskip 0.10cm
We then expect that the coupling among the whole basis functions
take place smoothly and coherently so as to
describe properly both the short-range structure and
long-range asymptotic behavior simultaneously.
We note that a single Gaussian  decays quickly as $x$ increases, 
but appropriate
superposition of many Gaussians can decay even  exponentially up to a
sufficiently large distance.  A good example is shown
in Fig.~3 of Ref.~\cite{Hiyama03} for the $^4$He dimer wave function
(with the HFDHE2 potential) 
that is accurate up to $\sim \! 1000$ \AA$\,$ 
with the use of the nonlinear parameters 
\{$n^{\rm max}=60$, $x_1=0.14 $ \AA~and $x_{n^{\rm max}}=700 $ \AA$\,$\}
(the same-quality dimer wave function is seen in 
Fig.~2 below in Sec.II.D using the complex-range Gaussians
with the LM2M2 potential).  
 
%\vskip 0.1cm
A lot of successful 
examples of the three- and four-body GEM calculations are shown 
in review papers~\cite{Hiyama03,Hiyama09,Hiyama10} 
and in papers of five-body calculations~\cite{Pentaquark,Hida}.
The examples includes our previous calculation of the ground  
and excited states of $^4$He trimer using the 
HFDHE2 potential;
the binding energies were  in good agreement with those
given by a Feddeev-equation calculation~\cite{Gloeckle86}.
As for the trimer wave function, we showed,
in Figs.~3, 13 and 14 in Ref.\cite{Hiyama03},  that  
the strong short-range correlation ($x \lap 4$ \AA) 
and asymptotic behavior (up to $ x \sim 1000 $ \AA)
of  the trimer ground and excited states
are simultaneously well described. 
Also, the three-body basis functions
(2.9)--(2.13) together with the LM2M2 potential were used recently
by Naidon, Ueda and one of the present authors (E. H.)~\cite{Naidon2011}
to study the universality and the three-body parameter of
$^4$He trimers.

%%%%%%%%%%%%%%%%%%%%%%%%%%%%%%%%%%%%%%%%%%%%%%%%%%%%%%%%
\subsection{Complex-range Gaussian basis functions}
%%%%%%%%%%%%%%%%%%%%%%%%%%%%%%%%%%%%%%%%%%%%%%%%%%%%%%%%

Before we proceed to the calculation of 
the $^4$He tetramer ground and excited states,
we improve the Gaussian shape of the basis functions
so as to have more sophisticated (but still 
tractable) radial dependence. 
We then test the new basis in the calculation of the trimer states
below.

In Ref.~\cite{Hiyama03}, we proposed to improve 
the Gaussian shape   
by introducing  {\it complex}  range 
instead of the real one: 
\begin{equation}
\phi_{nl}^{(\omega)}(x) =
x^l\:e^{- ( 1 + i\, \omega )(x/x_n)^2 } ,
\end{equation}
where $n=1, ..., n_x^{\rm max}$ and $x_n$ are given by (2.12).
Using % the complex conjugate pair
$\phi_{nl}^{( \pm \omega)}(x)$, we construct
two kinds of {\it  real} basis functions:
\begin{eqnarray}
\phi_{nl}^{({\rm cos})}(x) &=&  
x^l\:e^{-(x/x_n)^2}\, {\rm  cos}\, \omega (x/x_n)^2 \nonumber \\
& =&
[ \phi_{nl}^{(-\omega)}(x)+\phi_{nl}^{(\omega)}(x) ]/2, \\
\phi_{nl}^{({\rm sin})}(x) &=& 
x^l\:e^{-(x/x_n)^2}\, {\rm  sin}\, \omega (x/x_n)^2 \nonumber\\
& =&
[ \phi_{nl}^{(-\omega)}(x)-\phi_{nl}^{(\omega)}(x) ]/2i,
\end{eqnarray}
where we usually take $\omega=1$.
The three-body basis function 
$\Phi_\alpha( {\bf x}_i, {\bf y}_i)$ in (2.9)
is replaced by 
\begin{equation}
 \Phi_\alpha( {\bf x}_i, {\bf y}_i)=
\phi^{(^{\rm cos}_{\rm sin})}_{n_x l_x}(x)
% \phi_{n_x l_x}(x_i)
     \,   \psi_{n_y l_y}(y_i) 
\Big[ Y_{l_x}({\widehat {\bf x}_i}) %\otimes
Y_{l_y}({\widehat {\bf y}_i})
  \Big]_{J M},
\end{equation}
where $\alpha$  specifies a set 
\begin{equation}
\alpha \equiv \mbox{
 \{`cos' or `sin',} \, \omega,
n_x l_x, n_y l_y,JM \}.
\end{equation}

The new basis \{$\phi^{(^{\rm cos}_{\rm sin})}_{n_x l_x}(x)$\}
apparently extend the function space from 
the old ones (2.10) since they have the oscillating 
components; 
see Sec.2.4 and Sec.2.5 of Ref.~\cite{Hiyama03} 
for some examples taking this advantage  in calculations of
highly vibrational excited states (with $\sim$ 25 nodes) and 
scattering states.
The {\it sin}-type basis (2.17)
particularly work
when the wave function is extremely suppressed at $x \sim 0$
due to the strongly repulsive short-range potential.

%\vskip 0.1cm
In the following calculations,
we employ the new basis (2.16) and (2.17) for the $x$-space
instead of (2.10), but keep (2.11) for the $y$-space.

Note that, when calculating the matrix elements (2.7) and (2.8)
using $\phi_{nl}^{({\rm cos})}(x)$ and $\phi_{nl}^{({\rm sin})}(x)$,
we explicitly  take (2.15) and the right-most 
expression of (2.16) and (2.17)
since the computation programming 
is almost the same as that for (2.10) though some of real variables are 
changed to complex ones.

A great advantage of the  real- and complex-range Gaussian
basis  functions is that
the calculation of matrix elements (2.7) and (2.8)
is  easily performed.
As for the overlap and  kinetic-energy matrix elements
of the trimer (tetramer), all the six-(nine-)dimensional
integrals give analytical expression.
In the case of the potential matrix,
we have analytical expression
except for the one-dimensional numerical integral
having the final form
\begin{equation}
\int_0^\infty x^{2 m}\,e^{-\lambda x^2}\,V(x)\,x^2\,dx \, .
%\int_0^\infty x_i^{2\,m}\,e^{-\alpha\,x_i^2}\,V(x_i)\,x_i^2\,dx_i \,
%\quad (i=1 - 3) .
\end{equation}
We explained, in Ref.~\cite{Hiyama03}, various techniques to perform the
three- and four-body matrix-element calculations 
as easily, accurately and rapidly as possible.

%\vskip 0.1cm
It is to be emphasized that  the GEM few-body calculations 
need neither introduction of 
any {\it a priori} pair correlation function 
(such as the Jastrow function) 
nor separation of the coordinate space by $x< r_c $ and $x> r_c$, 
$r_c$ being the radius of a strongly repulsive core potential.  
Proper short-range correlation and  asymptotic
behavior of the total wave function 
are  {\it automatically} obtained
by solving the Schr\"{o}dinger equation (2.2)  
using the above  basis functions for {\it ab initio} calculations.

%%%%%%%%%%%%%%%%%%%%%%%%%%%%%%%%%%%%%%%%%%%%%%
\subsection{Pair interaction and $^4$He dimer}
%%%%%%%%%%%%%%%%%%%%%%%%%%%%%%%%%%%%%%%%%%%%%%

To describe the interaction between the $^4$He atoms, 
we employ one of the most widely used $^4$He-$^4$He interactions, 
the LM2M2  potential by Aziz and Slaman~\cite{LM2M2}.
Use is made of $\frac{\hbar^2}{m}=12.12$ K\AA$^2\,$ 
as the input mass of $^4$He atom.
We can then precisely compare calculated results for the 
tetramer ground and excited states
with those obtained by Lazauskas and Carbonell~\cite{Carbonell} 
who made a Faddeev-Yakubovsky (FY) equation calculation
taking the same potential and $^4$He mass as above.
Recently, the authors of Ref.~\cite{newmass} claim that
a more precise value of 
$\frac{\hbar^2}{m}=12.11928$ K\AA$^2\,$
should be employed.  We shall additionally show the trimer
and tetramer binding energies in the case of using this value.
%%%%%%%%%%%%%%%%%%%%%%%%%%%%%%%%%%%%%%%%%%%%%%
\begin{figure}[htb]
\begin{center}
%\begin{minipage}[t]{6.0 cm}
\epsfig{file=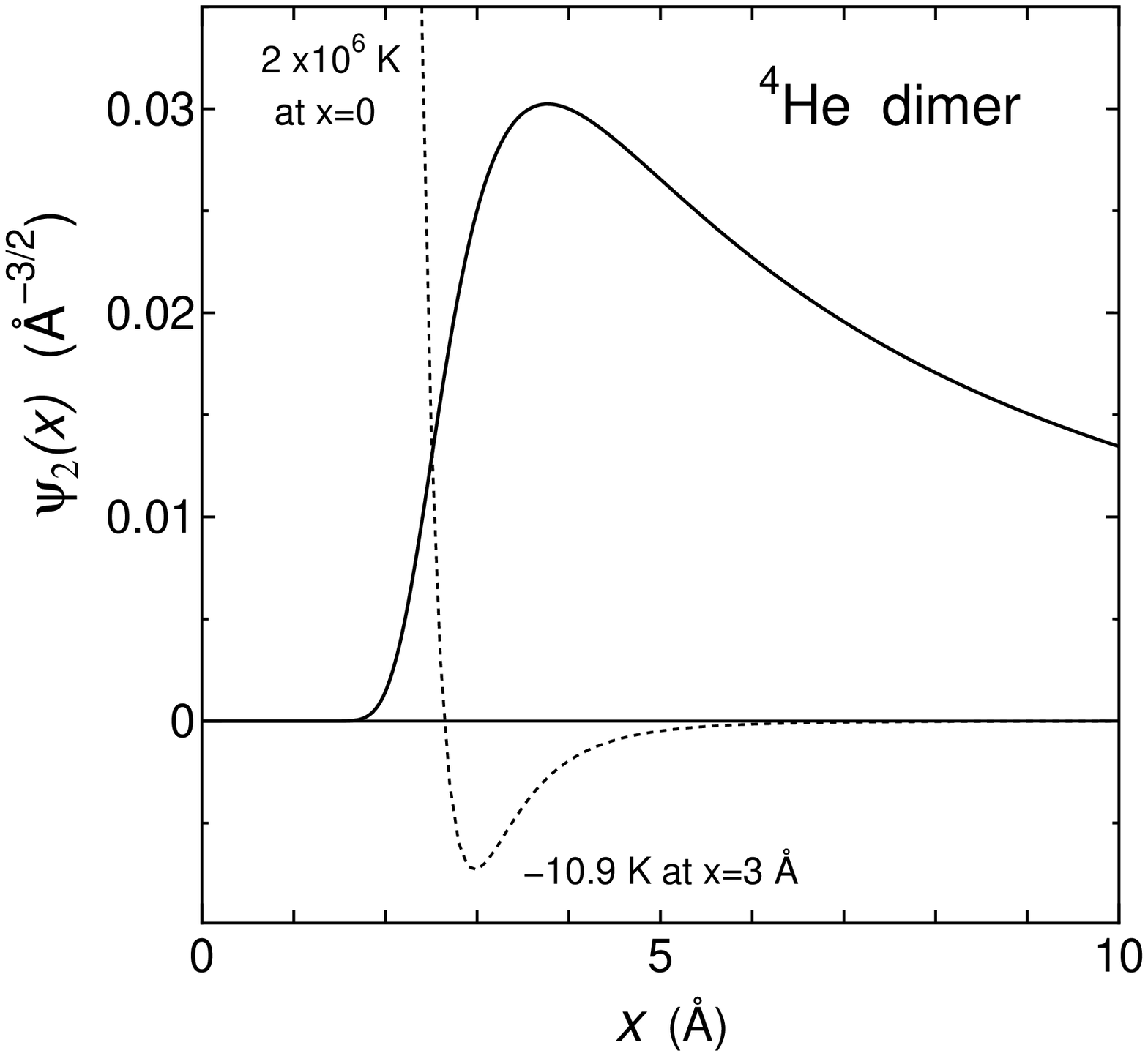,width=6.5cm,height=5.5cm}
%\end{minipage}
%\hspace{\fill}
%\begin{minipage}[t]{6.0 cm}
\epsfig{file=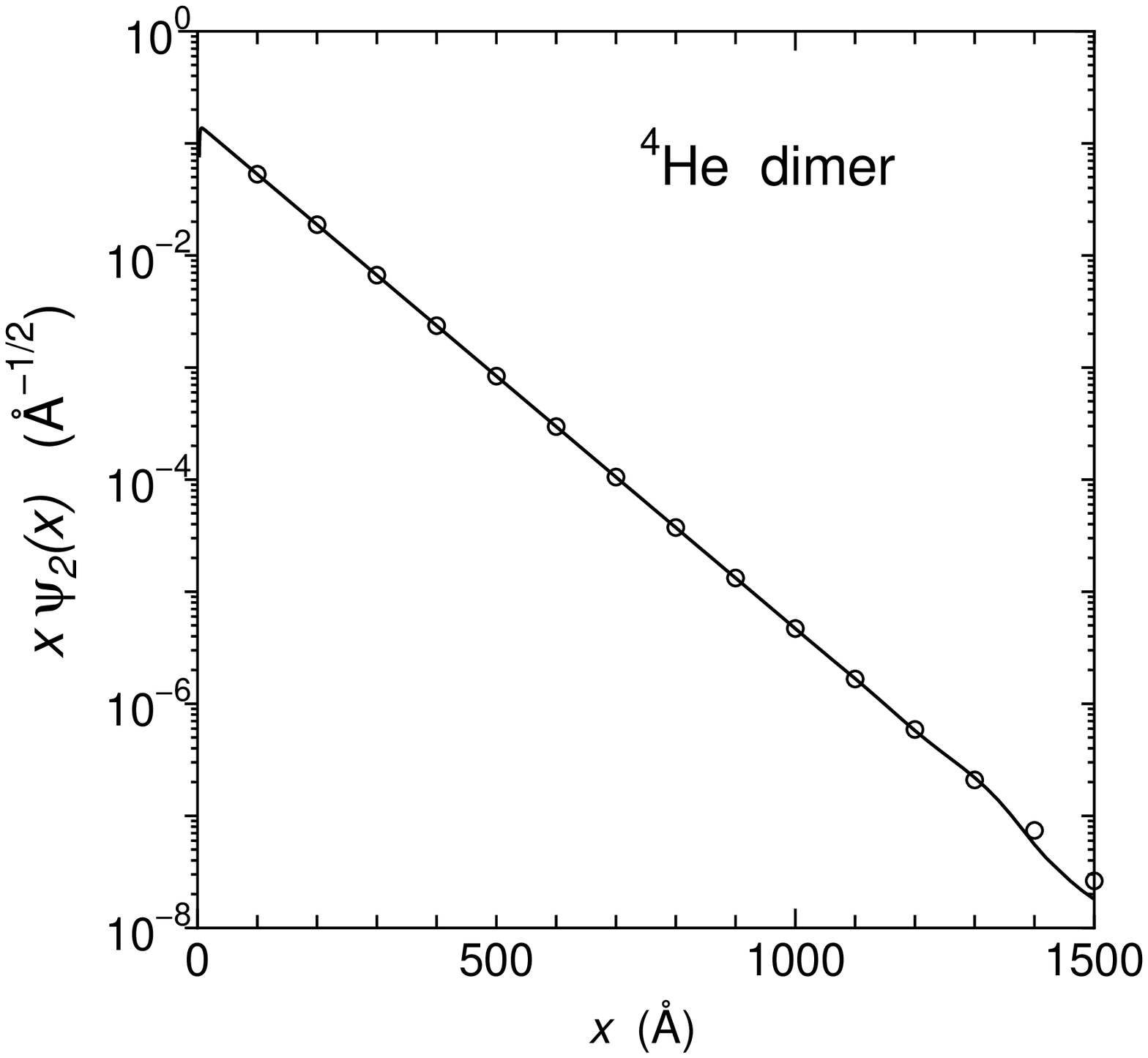,width=6.5cm,height=5.5cm}
%\end{minipage}
\end{center}	
\caption{
Short-range structure (upper) and asymptotic behavior (lower) 
of the radial wave  function $\Psi_2(x)$ of the $^4$He dimer
obtained by using the complex-range Gaussian basis 
functions (2.21) and (2.22).
The open circles stands for the exact 
asymptotic form.
The dotted line (upper) illustrates 
the LM2M2 potential in arbitrary unit.
}
\label{fig:dimer}
\end{figure}
%%%%%%%%%%%%%%%%%%%%%%%%%%%%%%%%%%%%%%%%%%%%%%

We calculated the $^4$He dimer binding energy, say $B_2$,
and the wave function, $\Psi_2 ( \equiv
\Psi_2(x) Y_{00}({\widehat {\bf x}})) $,
using the same prescription as described above.
We expanded $\Psi_2(x)$ with 100 basis functions of (2.16) and (2.17) as
\begin{equation}
\Psi_2(x) = \sum_{n=1}^{n_x^{\rm max}}
\big[ A_n^{\rm (cos)} \phi^{\rm (cos)}_{n 0} (x) 
      + A_n^{\rm (sin)} \phi^{\rm (sin)}_{n 0} (x) \big]
\end{equation}
with a parameter set 
\begin{equation}
\{n_x^{\rm max}=50, \,
x_1=0.5\, {\rm \AA}, \,  x_{n_x^{\rm max}}=600.0 
\,{\rm \AA}, \, \omega=1.0 \}.
\end{equation}
We obtained $B_2=1.30348$ mK, 
$\sqrt{\langle x^2 \rangle}=70.93$ \AA,
and $\langle x \rangle=52.00$ \AA$\,$which are the same as
those obtained in the literature.
Experimentally, 
$\langle x \rangle=52 \pm 4$ \AA~\cite{Grisenti2000}
from which  $B_2=1.1^{+0.3}_{-0.2}$ mK was estimated. 

%\vskip 0.1cm
As shown in Fig.~2,
both the strong short-range correlation  ($x \lap 4 $ \AA)
and the asymptotic behavior of the dimer are
well described. In the lower panel, 
$x\, \Psi_2(x)$ precisely reproduces the
exact asymptotic shape 
$0.1498 \,{\rm exp}(-\kappa_2 x)$ (\AA$^{-1/2}$) with 
$\kappa_2=\sqrt{m B_2}/\hbar=0.0104 $ \AA$^{-1}$
up to  $x \! \sim \!\! 1200 $ \AA $\,$ which is large enough for our
discussions. 

There are 30 basis functions whose Gaussian ranges $x_{n_x} < 4 $ \AA,
which is  sufficiently dense 
to describe  the short-range structure of
the wave function precisely.
An interesting issue is whether the same shape of the
short-range correlation 
in Fig.~\ref{fig:dimer}  appear also in
the trimer and tetramer ground and excited states
without assuming any two-body correlation function.

%%%%%%%%%%%%%%%%%%%%%%%%%%%%%%%%%%%%%%%%%%%%%%%
%%%%%%%%%%%%%%%%%%%%%%%%%%%%%%%%%%%%%%%%%%%%%%%
\subsection{Trimer bound states}
%%%%%%%%%%%%%%%%%%%%%%%%%%%%%%%%%%%%%%%%%%%%%%%

We calculated the wave functions of
the trimer ground state, $\Psi_3^{(0)}$, and the excited state,
$\Psi_3^{(1)}$, and their binding energies, $B_3^{(0)}$ and
$B_3^{(1)}$, respectively, as well as some mean values with 
the $\Psi_3^{(v)}$ $(v=0,1)$. 
Some of results  are summarized in Table I
together with those obtained in the literature.
Our results excellently agree  with those by Refs.\cite{Carbonell,
Kievsky01,Roudnev}.
The $^4$He-$^4$He bond length in the trimer ground state
was measured as 
$\langle r_{ij} \rangle = 11^{+4}_{-5}$ \AA~\cite{Grisenti2000},
which is well explained by the calculations,
$\langle r_{ij} \rangle=9.61$ \AA.  

%%%%%%%%%%%%%%%%%%%%  Table I %%%%%%%%%
\begin{table}[t]
\caption{ Mean values for $^4$He trimer ground and excited states
with the use of the LM2M2 potential and 
$\frac{\hbar^2}{m}=12.12$ K\AA$^2\,$.
$B_3^{(v)}$ is  the binding energy, $r_{ij}$ stands for interparticle
distance and $r_{i{\rm G}}$ is the distance of a particle from the
center-of-mass of the trimer. 
See text for the
asymptotic normalization coefficient $C_3^{(v)} (v=0,1)$.
}
\begin{center}
\begin{tabular}{crrrr} 
\hline \hline
\noalign{\vskip 0.1 true cm} 
trimer   & \multicolumn{4}{c} {ground state}  \\
\noalign{\vskip 0.1 true cm} 
\hline 
\noalign{\vskip 0.1 true cm} 
 & \quad present \quad & \quad Ref.\cite{Carbonell}
 \quad  &\quad Ref.\cite{Kievsky01} \quad 
& \quad Ref.\cite{Roudnev} \quad \\
\noalign{\vskip 0.1 true cm} 
\hline 
\noalign{\vskip 0.1 true cm} 
  $B_3^{(0)}$  (mK) 
&   126.40   &  126.39  &  126.4  &  126.40  \\
 $\langle T \rangle $  (mK)   
&  1660.4    &  1658  &   1660  &  \\
 $\langle V \rangle $  (mK)
&  $-1786.8$   &  $-1785$   & $-1787$   &  \\
 $ \sqrt{ \langle r^2_{ij} \rangle  }$ (\AA)
&  $10.96$   &  10.95   & 10.96   &  \\
 $  \langle r_{ij} \rangle  $ (\AA)
&  $9.616$   &  9.612   & 9.610   &  \\
 $  \langle r_{ij}^{-1} \rangle  $ (\AA$^{-1})$
&  $0.134$   &  0.135   &    &  \\
 $  \langle r_{ij}^{-2} \rangle  $ (\AA$^{-2})$
&  $0.0228$   &  0.0230   &     &  \\
 $ \sqrt{ \langle r_{i{\rm G}}^{2} \rangle } $ (\AA)
&  6.326   &     &  6.49 & 6.32 \\
\noalign{\vskip 0.1 true cm} 
  $C_3^{(0)}$(\AA$^{-\frac{1}{2}})$ 
&  0.562   &     &  0.567 &      \\
%%%%%%%%%%%%%%%%%%%%%%%%%%%%%%%%%%%%%%%%%%%%%%%%%%%%%%%%%%%%%%%%%%%%%%
\noalign{\vskip 0.1 true cm} 
\hline 
\noalign{\vskip 0.2 true cm} 
trimer  &    \multicolumn{4}{c}  {excited state}   \\
\noalign{\vskip 0.1 true cm} 
\hline 
\noalign{\vskip 0.1 true cm} 
 & \quad present \quad & \quad Ref.\cite{Carbonell}
 \quad  &\quad Ref.\cite{Kievsky01} \quad 
& \quad Ref.\cite{Roudnev} \quad \\
\noalign{\vskip 0.1 true cm} 
\hline 
\noalign{\vskip 0.1 true cm} 
  $B_3^{(1)}$  (mK) 
& 2.2706  & 2.268   & 2.265  & 2.2707   \\
 $\langle T \rangle $  (mK)   
&  122.15  &  122.1   & 121.9 &    \\
 $\langle V \rangle $  (mK)   
&  $-124.42$   &  $-124.5$   &   $-124.2$   &  \\
 $ \sqrt{ \langle r^2_{ij} \rangle  }$ (\AA)
&  $104.5$   &  104.3   & 101.9  &    \\
 $  \langle r_{ij} \rangle  $ (\AA)
&  84.51   &  83.53    & 83.08 &   \\
 $  \langle r_{ij}^{-1} \rangle  $ (\AA$^{-1})$
&  0.0265   &  0.0267   &  &    \\
 $  \langle r_{ij}^{-2} \rangle  $ (\AA$^{-2})$
&  0.00216   &  0.00218   &  &  \\
 $ \sqrt{ \langle r_{i{\rm G}}^{2} \rangle } $ (\AA)
&  60.33   &    & 58.8   &  59.3  \\
\noalign{\vskip 0.1 true cm} 
   $C_3^{(1)}$(\AA$^{-\frac{1}{2}})$   %% {-1/2}$)
&  0.179   &     &  0.178 &      \\
\noalign{\vskip 0.2 true cm} 
\hline \hline
\end{tabular}
\label{table:mean-trimer}
\end{center}
\end{table}
%%%%%%%%%%%%%%%%%%%%%%%%%%%%%%%%%%%%%%

Those converged results were given by taking the symmetric
three-body basis function 
\{$\Phi_\alpha^{\rm (sym)}; \alpha=1, ..., \alpha_{\rm max}$\}
with $\alpha_{\rm max}=4400$, in which the shortest-range
set is $(x_1\!=\!0.3\, $\AA,$\,\, y_1\!=\!0.4\, $\AA) 
and the longest-range one is 
$(x_{\rm max}\!=\!150\, $\AA,$\,\, y_{\rm max}\!=\!600 \,$\AA). 
All the nonlinear  parameters of the Gaussian basis set
are listed in \mbox{Table II}.

%%%%%%%%%%%%%%%%%  Table II   %%%%%%%%%%%%
\begin{table}[t]
\caption{ All the nonlinear parameters of the Gaussian basis functions
used for the $^4$He-trimer states with $J=0 \, (l_x=l_y$).  
Those in column a) are commonly for 
$\phi^{({\rm cos})}_{n_x l_x}(x)$ and $\phi^{({\rm sin})}_{n_x l_x}(x)$
and b) for $\psi_{n_y l_y}(y)$. Total number of the basis, 
$\alpha_{\rm max}$, is 4400.
}
\begin{center}
\begin{tabular}{ccccccccccc}
\hline
\hline
\noalign{\vskip 0.2 true cm} 
   \multicolumn{5}{c} 
{ a) $\phi^{({\rm cos})}_{n_x l_x}(x),\, \phi^{({\rm sin})}_{n_x l_x}(x)$ } 
&$\;\;\;$ &  \multicolumn{4}{c} { b) $\psi_{n_y l_y}(y)$} &  \\
\noalign{\vskip 0.1 true cm} 
   \multispan5 {\hrulefill} & \qquad &   \multispan4 {\hrulefill} & \\
\noalign{\vskip 0.1 true cm} 
 $l_x $ & $n_x^{\rm max} $ & $x_1$ & $x_{n_x^{\rm max}}$ & $\omega$
& \qquad & $l_y$ & $n_y^{\rm max} $ & $y_1$  & $y_{n_y^{\rm max}}$ 
& $\;$ number\\
%\vspace{-3 mm} \\
%\noalign{\vskip 0.1 true cm} 
        &              &      [\AA] &             [\AA]  &
& \qquad &       &      &       [\AA] &            [\AA] 
& $\;$ of basis\\
%\vspace{-3 mm} \\
\noalign{\vskip 0.1 true cm} 
\hline
\vspace{-3 mm} \\
 0   &  22  &  0.3 &   150.0& 1.0 & \qquad  & 0 & 50  &  0.4 &   600.0
  & 2200 \\
 2   &  17  &  0.6 &   150.0& 1.0 &  \qquad & 2 & 40  &  0.8 &   400.0 
  & 1360 \\
 4   &  14  &  0.8 &   130.0&  1.0 & \qquad & 4 & 30  &  1.0 &   200.0 
  & $\;\,$840\\
\vspace{-3 mm} \\
\hline
\hline
\end{tabular}
\end{center}
\end{table}
%%%%%%%%%%%%%%%%%%%%%%%%%%%%%%%%%%%%

There are neither additional parameter nor assumptions.
The present calculation is so transparent that
it is possible for the readers to repeat the calculation
and check the results reported here.
The parameters for the Gaussian ranges are in  round numbers
but further optimization of them do not improve the 
binding energies ($B_3^{(0)}=126.40$ mK and $B_3^{(1)}=2.2706$ mK)
as long as we calculate them with  {\it five}
significant figures (cf. another check in Sec.II.H about the accuracy 
of the calculation).

Convergence of the binding energies $B_3^{(0)}$ and $B_3^{(1)}$ 
with respect to increasing
partial waves $l_x (=l_y)$ 
is shown in Table III in comparison with the Faddeev
calculation by Lazauskas and Carbonell~\cite{Carbonell}.
The case  $l_{x_{\rm max}}=4$ is sufficient in the present  work
as long as the accuracy of five significant digits is required.

The convergence of the present result is
more rapid than that of the Faddeev solution (the same will
be seen in the tetramer calculation in Sec.III).
The reason is that both the interaction and the wave function 
are truncated in the angular-momentum space $(l_{\rm max})$
in the Faddeev calculations,
but the full interaction
is included in the present calculation 
(with no partial-wave decomposition) though
the wave function is truncated  $(l_{\rm max})$.
The difference of the convergence in the
two calculation methods was precisely
discussed in the case of the three nucleon 
bound states  ($^3$H and $^3$He nuclei)
in our GEM calculation~\cite{Kameyama89,Kamimura90,Hiyama03}  
and in a Faddeev calculation~\cite{Payne93}; 
for an illustration of 
the difference, see Fig.~15 in Ref.~\cite{Hiyama03}. 
In this context, it is worth pointing out that, in 
Table~\ref{table:mean-trimer},
our result  precisely agrees with another Faddeev 
calculation by Ref.\cite{Roudnev} with no
the partial wave decomposition.

%%%%%%%%%%%%%%%    Table III  %%%%%%%%%%%%%%
\begin{table}[h]
\caption{ Convergence of the $^4$He trimer calculations
with respect to the increasing maximum partial wave $(l_{\rm max})$.
The four columns present
trimer ground $(B_3^{(0)})$ and excited $(B_3^{(1)})$ state energies
in comparison with those obtained by the Faddeev-equation 
calculation of Ref.\cite{Carbonell}.
}
\label{table:1}
\begin{center}
\begin{tabular}{cccccc} 
\hline \hline
\noalign{\vskip 0.1 true cm} 
 trimer  & \multicolumn{2}{c} {present}  & \qquad $\;$
  &  \multicolumn{2}{c}  {Ref.\cite{Carbonell}}   \\
\noalign{\vskip -0.2 true cm} 
 &  \multispan2 {\hrulefill} & & \multispan2 {\hrulefill} \\
\noalign{\vskip 0.1 true cm} 
 $l_{\rm max}$   
&  $B_3^{(0)}$ (mK)   & $B_3^{(1)}$ (mK) & & 
$B_3^{(0)}$ (mK)  & $B_3^{(1)}$ (mK) \\
\noalign{\vskip 0.2 true cm} 
\hline 
\noalign{\vskip 0.2 true cm} 
  0   &   121.00   &   2.2397  &  &  89.01        &  2.0093 \\
  2   &  126.39   &   2.2705   &  &  120.67          &  2.2298 \\
  4    &  126.40    &  2.2706  &  &  125.48   & 2.2622 \\
  8   &    &          &  &  126.34    & 2.2677 \\
 12   &     &         &   &  126.39   & 2.2680 \\
 14   &     &         &   &  126.39   & 2.2680 \\
%\vspace{-5 mm} \\
\noalign{\vskip 0.1 true cm} 
\hline
\hline
\end{tabular}
\label{table:resonance}
\end{center}
\end{table}
%%%%%%%%%%%%%%%%%%%%%%%%%%%%%%%%%%%%%%

Use of the value 
$\frac{\hbar^2}{m}=12.11928$ K\AA$^2\,$~\cite{newmass} 
results in  $B_3^{(0)}=126.499$  mK 
and $B_3^{(1)}=2.27787$  mK,  while Ref.~\cite{newmass}
gives $126.499$  mK and $2.27844$  mK, respectively.
Calculation of  the binding  energy was also made  perturbatively 
with $B_3= \frac{12.11928}{12.12} \langle \, T \, \rangle
+ \langle \, V \, \rangle$, where 
$ \langle \, T \, \rangle$ and $ \langle \, V \, \rangle$
are those obtained with $\frac{\hbar^2}{m}=12.12$ K\AA$^2$;
this gives $B_3^{(0)}=126.498$ mK and
$B_3^{(1)}=2.27787$ mK. 
The calculations below in Sec.II.F-H take 
$\frac{\hbar^2}{m}=12.12$ K\AA$^2$.

%%%%%%%%%%%%%%%%%%%%%%%%%%%%%%%%%%%%%%%%%%%%%%%
%%%%%%%%%%%%%%%%%%%%%%%%%%%%%%%%%%%%%%%%%%%%%%%
\subsection{Short-range correlation and asymptotic behavior}
%%%%%%%%%%%%%%%%%%%%%%%%%%%%%%%%%%%%%%%%%%%%%%%

In order to see how the present method 
describes the short-range structure of  trimer,
we calculated the  pair correlation function 
(pair distribution function or two-body density)
$P_3^{(v)}(x)$ defined by
\begin{equation}
P_3^{(v)}(x_1)\, Y_{00}({\widehat {\bf x}}_1) = 
\langle \,\Psi_3^{(v)}\,\!|\,\Psi_3^{(v)}\,
\rangle_{{\bf y}_1} ,
\end{equation}
where the symbol $\langle \quad \rangle_{{\bf y}_1}$ means
the integration over ${\bf y}_1$ only.
This integration gives an analytical expression
owing to the use of the Gaussian basis functions; here, we explicitly
rewrite $\Psi_3^{(v)}$ as a function of  $({\bf x}_1, {\bf y}_1)$ 
by transforming the other coordinates 
$({\bf x}_2, {\bf y}_2)$ and $({\bf x}_3, {\bf y}_3)$ 
into $({\bf x}_1, {\bf y}_1)$.
$P_3^{(v)}(x_i)$ is independent of $i \, (=1, 2,  3)$
and is apparently normalized as 
$\int P_3^{(v)}(x)\, x^2 {\rm d}x=1$. It
presents the probability of finding two particles at
an interparticle distance $x$.

\begin{figure}[h]
\begin{center}
\epsfig{file=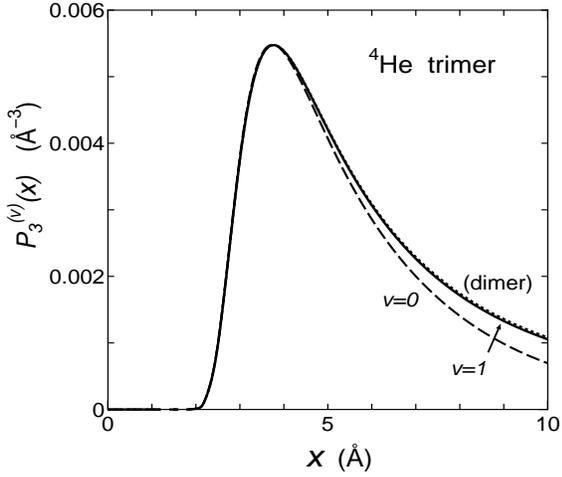,width=7.5cm,height=6.4cm}
\caption{
Short-range structure of the pair correlation function
$P_3^{(v)}(x)$ of the $^4$He trimer calculated by (2.23). 
The dashed line is for the trimer
ground state $(v=0)$, the solid line for the excited state $(v=1)$
and the dotted line for the $^4$He dimer ($|\Psi_2(x)|^2$). 
The solid and  dotted lines have been multiplied by  
factors 14.5 and 6.0, respectively, to be normalized at the peak.
The same shape of the short-range correlation 
($x \lesssim 4 $ \AA $\,$) appears in the three states.
}
\label{fig:trimer-den-short}
\end{center}
\end{figure}

In Fig.~\ref{fig:trimer-den-short}, 
short-range structure of $P_3^{(v)}(x) (v=0,1)$
is illustrated together with $P_2(x)(= |\Psi_2(x)|^2)$ 
for the $^4$He dimer.
The dashed line is for the trimer ground state $(v=0)$.
The solid line for the excited state $(v=1)$
and the dotted line for the  dimer 
have been multiplied by  factors 14.5 and 6.0, respectively.
It is of interest that 
precisely the same shape of the short-range 
correlation $(x \lap 4 $ \AA) as seen in the dimer
appears both in the trimer ground and excited states
(the same will be seen in the tetramer
ground and excited states in the next section). 
This gives a foundation to
an {\it a priori} assumption that a two-particle 
correlation function (such as the Jastrow function) 
to simulate the short-range part of
the dimer radial wave function $\Psi_2(x)$ 
is incorporated in the three-body  
wave function from the beginning.

%%%%%%%%%%%%%%%%%%%%%%%%%%%%  Fig. 3   %%%%%%%%%%%%%%%%%%
%%%%%%%%%%%%%%%%%%%%%%%%%%%%

\begin{figure}[t]
\begin{center}
\epsfig{file=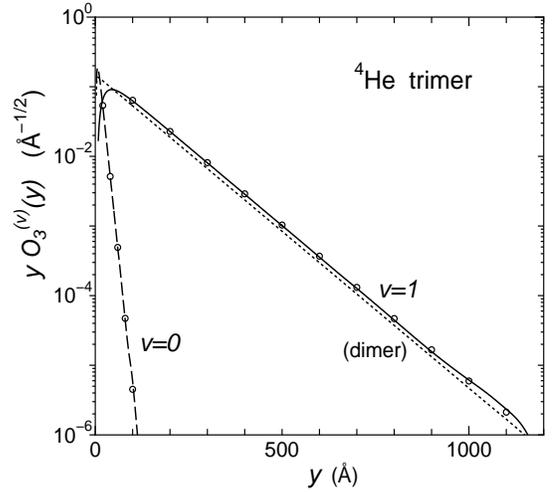,width=7.2cm,height=6.6cm}
\caption{
Overlap function ${\cal O}_3^{(v)}(y)$, multiplied by $y$,
between the $^4$He trimer wave function $(v=0,1)$
and the dimer one, which is defined by (2.24).
Open circles represent the fit of the 
asymptotic function (2.25) to ${\cal O}_3^{(v)}(y)$
using the asymptotic normalization coefficient
$C_3^{(v)}$.
The solid line $(v=1)$ is found to be parallel to
the dotted line for the dimer wave function ($y\Psi_2(y)$). 
}
\label{fig:trimer-red-long}
\end{center}
\end{figure}

%\vskip 0.1cm
To investigate the trimer configuration
in the asymptotic region where
one atom is far from the other two,
we calculate the overlap function 
${\cal O}_3^{(v)}(y_1)$~\cite{Friar,Kameyama89,Kievsky01,ANC1,ANC2}
to describe the overlap between 
the trimer wave function $\Psi_3^{(v)} (v=0,1)$ and 
the dimer one $\Psi_2({\bf x})$:
\begin{equation}
{\cal O}_3^{(v)}(y_1)\, Y_{00}({\widehat {\bf y}}_1)  =  
\langle \, \Psi_2({\bf x}_1) \, | \,
\Psi_3^{(v)} \, \rangle_{{\bf x}_1}.
\end{equation}
In Fig.~\ref{fig:trimer-red-long}, we plot $y \, {\cal O}_3^{(v)}(y)$ 
for the ground  and excited  states.
They should asymptotically satisfy
\begin{eqnarray}
y \,{\cal O}_3^{(v)}(y) \stackrel{y \to \infty}{-\!\!\!\longrightarrow}
 C_3^{(v)}{\rm exp}(-\kappa_3^{(v)} y), 
\end{eqnarray}
where $\kappa_3^{(v)}$ is the binding wave number given by
$\kappa_3^{(v)}\!=\!\sqrt{2\mu_y (B_3^{(v)}-B_2)}/\hbar$ 
($\kappa_3^{(0)}\!=\!0.117\!$~\AA$^{-1}$,
$\kappa_3^{(1)}\!=\!0.0103\!$~\AA$^{-1}$). The amplitude 
 $C_3^{(v)}$ is 
called the asymptotic normalzation 
coefficient (ANC)~\cite{Friar,Kameyama89,Kievsky01,ANC1,ANC2}
defining the amplitude of the tail of the radial overlap function.
%which reflects the internal three-body structural information.
The asymptotic functions (2.25) with
$C_3^{(0)}\!=0.562 $\AA$^{-\frac{1}{2}}$ and
$C_3^{(1)}\!=0.179 $\AA$^{-\frac{1}{2}}$ (see the open circles)
are precisely reproduced by the dashed line ($v=0$) and
the solid line ($v=1$), respectively, 
up to $y\! \sim\! 1000$ \AA$\,$,
which demonstrates the accuracy of 
our wave functions in the asymptotic region.
The values of $C_3^{(v)}$ agree with those given by 
Barletta and Kievsky~\cite{Kievsky01} using a variational method
with correlated hyperspherical harmonics functions (see Table I).

%%%%%%%%%%%%%%%%%%  Fig. 4 ( reduce trimer)  %%%%%%%%%%%%%
%%%%%%%%%%%%%%%%%%%%%%%%%%%%%%%%%%%%%%%

%\vskip 0.1cm
The total three-body wave function $\Psi_3^{(v)} (v=0,1)$ is represented
asymptotically  as 
\begin{eqnarray}
\Psi_3^{(v)} %% \stackrel{y_i \to \infty}{-\!\!\!-\!\!\!\longrightarrow} 
\longrightarrow C_3^{(v)} \sum_{i=1}^3
\Psi_2({\bf x}_i)
\frac{e^{-\kappa_3^{(v)} y_i}}{y_i}
Y_{00}({\widehat {\bf y}_i}) .  \quad
\end{eqnarray}
The ANC, $C_3^{(v)}$,
is a quantity to convey the interior structural information 
of the trimer to the asymptotic behavior.
It is known, in the nuclear peripheral reactions where 
only the asymptotic tails of the wave functions of
reacting particles contribute 
to the reaction process, the cross section is
proportional to the squared ANC which can be measured in some 
specific systems~\cite{ANC1,ANC2,Ogata-ANC-2006}. 
The idea of ANC might be available
to the calculation of $^4$He atoms reactions
such as ${\rm dimer} + {\rm dimer} \to {\rm trimer} + {\rm atom}$. 
Also, attention to the ANC might be useful
when one intends to  reproduce the non-universal variations
of the trimer states by parametrizing 
effective models.

%\vskip -0.3cm
%%%%%%%%%%%%%%%%%%%%%%%%%%%%%%%%%%%%%%%%%%%%%%%%%%%%%%
\subsection{'Dimerlike-pair' model in asymptotic region}
\begin{figure}[t]
\begin{center}
\epsfig{file=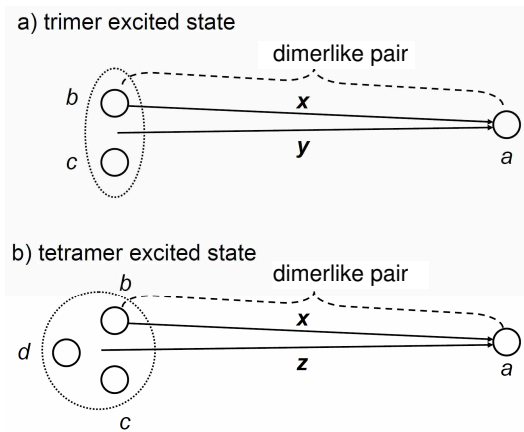,width=220pt}
\end{center}
\caption{
'Dimerlike-pair' model for the asymptotic behavior 
of the trimer and tetramer excited states 
(see text).
}
\label{fig:dimer-model}
\end{figure}

%%%%%%%%%%%%%%%%%%%%%%%%%%%%%%%%%%%%%%%%%%%%%%%%%%%%%%
%\vskip -0.3cm
In Fig.~\ref{fig:trimer-red-long}, we find that
the solid line ($v=1$) is parallel to the dotted line (dimer);
namely, $\kappa_3^{(1)}(=0.0103$ \AA$^{-1}$) is very close to
$\kappa_2(=0.0104$ \AA$^{-1}$).
This agreement is not accidental, but is
understandable from a model, which we refer to as a 'dimerlike-pair' model, 
for the asymptotic behavior of the trimer excited state
(Fig.~\ref{fig:dimer-model}a).
The model tells that 
i) particle $a$, located far from 
$b$ and $c$ which are {\it loosely} bound (dimer), is little affected
by the interaction between $b$ and $c$,
ii) therefore, the pair $a$ and $b$ at a distance $x$ 
is asymptotically dimerlike,
iii) since \mbox{{\bf x} $\simeq$ {\bf y}}  asymptotically, 
the amplitude of particle $a$ 
along {\bf y} is dimerlike, namely  
$\kappa_3^{(1)} \simeq \kappa_2$.

%%%%%%%%%%%%%%%%%%  Fig. 5 (dimer-model) %%%%%%%%%%%
%%%%%%%%%%%%%%%%%%%%%%%%%%%%%%%%%%%%%%%%%%%%%%%%%

\begin{figure}[t]
\begin{center}
\epsfig{file=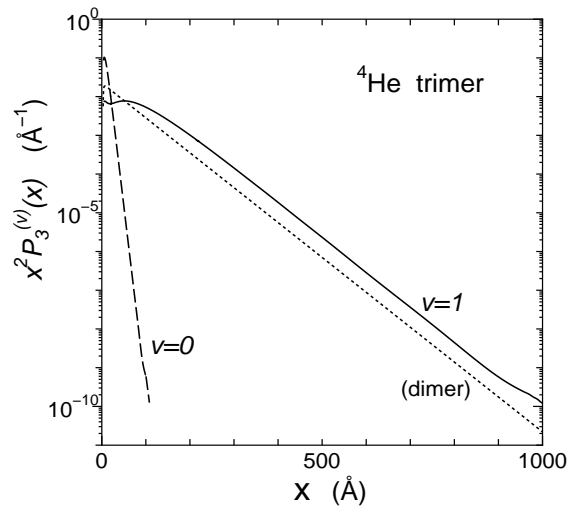,width=7.5cm,height=6.8cm}
\caption{
Asymptotic behavior of the pair correlation (distribution) function
$P_3^{(v)}(x)$, multiplied by $x^2$, of the $^4$He trimer.
The dashed line is for the trimer
ground state $(v=0)$. The solid line for the excited state $(v=1)$
and the dotted line for the dimer are found to have
almost the same exponentially-decaying rate, 
$2\kappa_3^{(1)} \simeq 2\kappa_2$.
}
\label{fig:trimer-den-long}
\end{center}
\end{figure}

If this model is acceptable, we can predict that, 
in the asymptotic region,
the pair correlation function of the trimer excited state, 
$x^2 P_3^{(1)}(x)$, should decay exponentially with the 
same rate as that in the dimer ($x^2 P_2(x)$).
This is clearly seen 
in Fig.~\ref{fig:trimer-den-long}; the solid and dotted lines
have almost 
the same exponentially-decaying rate 
of $2 \kappa_3^{(1)}( \simeq 2 \kappa_2$). 
The same evidence  is seen in Figs.~3 and 14 
of our previous calculation of the $^4$He trimer using the HFDHE2
potential reported in Ref.~\cite{Hiyama03}. 

Once we accept the dimerlike-pair model
($\kappa_3^{(1)} \simeq \kappa_2$),
we can estimate 
$B_3^{(1)}$, the trimer excited-state binding energy,
using $B_2$.  With the use of the definitions
of the binding wave numbers:
\begin{eqnarray}
\kappa_2 & = & \sqrt{2\mu_x B_2}/\hbar , \\
\kappa_3^{(1)} & = & \sqrt{2\mu_y (B_3^{(1)} - B_2)}/\hbar , 
\end{eqnarray}
where  $\mu_x=\frac{1}{2}m$ and $\mu_y=\frac{2}{3}m$. 
Taking $\kappa_3^{(1)} \simeq \kappa_2$, we can then predict  
\begin{equation}
 B_3^{(1)}  \simeq B_2 + \frac{\mu_x}{\mu_y} B_2
= \frac{7}{4} B_2 =2.281 \, {\rm mK},
\end{equation}
which is close to 2.2706 mK by  the
present three-body calculation using the LM2M2 potential
for which we have the ratio $B_3^{(1)}/B_2=1.74 \, (\simeq 7/4)$. 

In order to see a deviation of the ratio from 7/4
depending on the realistic potentials in the literature,
we refer to $B_3^{(1)}/B_2 =$ 
1.59~\cite{Kievsky01} (SAPT2~\cite{SAPT2}),
1.65~\cite{Suno08,newmass} (SAPT2007~\cite{SAPT2007}), 
1.74~\cite{Kievsky01,Roudnev} (TTY~\cite{TTY}),
1.74 (LM2M2), 
2.01~\cite{Motovilov01} (HFDHE2). 
The dimerlike-pair model provides a reason why the ratio
$B_3^{(1)}/B_2$ is located  around 7/4 in a narrow region of 
1.6--2.0.

%%%%%%%%%%%%%%%%%%%%%%%%  Fig. 6 %%%%%%%%%%%%%
%%%%%%%%%%%%%%%%%%%%%%%%%%%%%%%%%%%%%%%%%%%%%%

  We note that this model should be considered under the condition
that the $^4$He atoms are interacting with 
a realistic pair potential and should not be discussed  
in any situation where a large deformation of 
the strength is posed to the potential (cf. a discussion in Sec.III of 
Ref.~\cite{Braaten03} on the Efimov states in $^4$He trimer).

We try to apply the same model to the tetramer
excited state (Fig.~\ref{fig:dimer-model}b) and predict 
its binding energy $B_4^{(1)}$.   %%   using $B_2$ only. 
% with respect to the atom-trimer threshold ($B_3^{(0)}$).
Asymptotically, 
particle $a$  decays from the trimer $(b+c+d$) as
${\rm exp}(-\kappa_4^{(1)} z)$ with
\begin{equation}
\kappa_4^{(1)}  =  \sqrt{2\mu_z (B_4^{(1)} - B_3^{(0)})}/\hbar , 
\end{equation}
where  $\mu_z=\frac{3}{4}m$ is the reduce atom-trimer mass.
Taking $\kappa_4^{(1)} \simeq \kappa_2$, 
we predict $B_4^{(1)} $ as    % $B_2$ only:
\begin{equation}
B_4^{(1)}   \simeq  B_3^{(0)} +  \frac{\mu_x}{\mu_z}B_2
=  B_3^{(0)} + \frac{2}{3} B_2 = 127.27 \,{\rm mK} 
\end{equation}
when employing the calculated values of
$B_3^{(0)} $ and $B_2$ with  LM2M2.
In Sec.III, we  make a four-body  calculation of
the tetramer  with  LM2M2 and check the above prediction of 
$B_4^{(1)}$.

%
%%%%%%%%%%%%%%%%%%%%%%%%%%%%%%%%%%%%%%%%%%%%%%%
%%%%%%%%%%%%%%%%%%%%%%%%%%%%%%%%%%%%%%%%%%%%%%%
\subsection{Generalized eigenvalue problem}
%%%%%%%%%%%%%%%%%%%%%%%%%%%%%%%%%%%%%%%%%%%%%%%

In this subsection, we discuss about 
a technical subject on  a numerical trouble
which arises when solving the generalized eigenvalue problem (2.6).
This is due to the fact that 
the overlap matrix  ${\cal N}$ becomes almost  singular
when  a very large number of
nonorthogonal basis  functions $\{\Phi_{\alpha}^{\rm (sym)}$\} 
employed. In this case, because of the 
non-negligible round-off error in double-precision 
computation ($\simeq$16 decimal digits), 
we may obtain no solution of (2.6) or 
a solution that includes
some unphysically too deep erroneous bound state.
In order to overcome this trouble,
we took the following two steps:

%\vskip 0.15cm
{\it Step} i): we first 
diagonalize the overlap matrix ${\cal N}$:
%of (2.8) as 
\begin{equation}
\sum_{\alpha'=1}^{\alpha_{\rm max}} 
{\cal N}_{\alpha, \alpha'} C_{\alpha'}^{(N)} 
= \nu_N C_\alpha^{(N)},  
\end{equation}
where $ \alpha, N = 1, ..., \alpha_{{\rm max}}$. The eigenvalues 
$\nu_N$ are positive definite since
${\cal N}_{\alpha, \alpha'}={\cal N}_{\alpha',\alpha}$.
We then define a new, symmetrized
{\it orthonormal} basis set:
\begin{eqnarray}
\!\!\!\!\!\!\!\!\!\!\!\!&&{\widehat \Phi}_N^{\rm (sym)} 
=\frac{1}{\sqrt{\nu_N}}
 \sum_{\alpha=1}^{\alpha_{\rm max}}  C_\alpha^{(N)} 
\Phi^{\rm (sym)}_\alpha, \\ 
\!\!\!\!\!\!\!\!\!\!\!\!&&   \langle {\widehat \Phi}_N^{\rm (sym)} 
 \: | \: {\widehat \Phi}_{N'}^{\rm (sym)}  \rangle 
  =  \delta_{N, N'},
\end{eqnarray}
where $N,N'=1, ..., \alpha_{\rm max}$.
The generalized eigenvalue problem (2.6) are then equivalently
converted into a standard eigenvalue problem:
\begin{equation}
 \sum_{N'=1}^{\alpha_{\rm max}} \big[ {\widehat {\cal H}}_{N, N'} -
 E\, {\delta}_{N, N'} \big] \, {\widehat A}_{N'} = 0,
\end{equation}
where $  N = 1, ...,  \alpha_{{\rm max}}$, and
\begin{equation}
{\widehat {\cal H}}_{N, N'} =  
\langle \: {\widehat \Phi}_{N}^{\rm (sym)}  \:| \:H\: |\: 
{\widehat \Phi}_{N'}^{\rm (sym)} \: \rangle .
\end{equation}
Here, we arrange 
$\{{\widehat \Phi}_N^{\rm (sym)}\}$ 
in the decreasing order of $\nu_N$:
\begin{equation}
\nu_1 > \nu_2 > \cdot\cdot\cdot > \nu_N >
\cdot\cdot\cdot > \nu_{\alpha_{\rm max}}.
\end{equation}
When the nonorthogonality among the basis functions 
\{$\Phi_\alpha^{\rm (sym)}$\} 
is very large, some of
$\nu_N$ become extremely small and therefore
the large factor $1/\sqrt{\nu_N}$ may cause a serious cancellation
in the summation in (2.33).
Since the present calculation is performed by double-precision
computation,
such a large cancellation
may generate a substantial round-off error 
in (2.33)  and hence in the matrix elements (2.36).
This may  give rise to some  erroneous eigenstates in  (2.35)
that have unphysically huge binding energies.

%%%%%%%%%%%%%%%%%%%%%%%%%%%%
\begin{table}[t]
\caption{Stability of the calculated trimer binding energies 
of the lowest-lying two states against 
decreasing number (${\widehat N}_{\rm max}$) 
of symmetrized orthonormal three-body basis functions 
\{${\widehat \Phi}_N^{\rm (sym)}$\} of Eq.(2.33).
This assures accuracy of the conclusion
$B_3^{(0)}\!=\!126.40$ mK and $B_3^{(1)}\!=\!2.2706$ mK 
in Table~ I. 
}
\label{table:result}
\begin{center}
\begin{tabular}{cccc}
\hline
\hline
\vspace{-3 mm} \\
  $ {\widehat N}_{\rm max}$ & $(\Delta {\widehat N}_{\rm max}) $
 &   $B_3^{(0)}$ (mK)  & $B_3^{(1)}$ (mK)  \\
\vspace{-3 mm} \\
\hline
\vspace{-3 mm} \\
 3250 &  $  -$  &  $\quad 126.3999 \quad$ & $ 2.270606 $\quad   \\
 3240 & $( -10)$&  $126.3998 $ & $ 2.270605 $   \\
 3200 & $( -50)$&  $126.3995 $ & $ 2.270602 $   \\
 3150 & $(-100)$&  $126.3991 $ & $ 2.270594 $   \\
 2950 & $(-300)$&  $126.3975 $ & $ 2.270533 $   \\
 2750 & $(-500)$&  $126.3954 $ & $ 2.270484 $   \\
 2250 & $(-1000)$& $126.3657 $ & $ 2.270163 $   \\
%1750 & $(-1500)$& $125.8863 $ & $ 2.264822 $   \\
\vspace{-3 mm} \\
\hline
\hline
\end{tabular}
\end{center}
\end{table}
%%%%%%%%%%%%%%%%%%%%%%%%%%%%%%%%%%%%

%\vskip 0.10cm
{\it Step} ii): We therefore omit such members of
\{${\widehat \Phi}_N^{\rm (sym)}$\}
that have too small $\nu_N$.
The binding energies of such unphysical states
decreases quickly as the basis size % ${\widehat \alpha}_{\rm max}$ 
is reduced.  Finally, 
we reach an appropriate size, say ${\widehat N}_{\rm max}$, 
of the basis \{${\widehat \Phi}_N^{\rm (sym)}$\} for which
those unphysical states have disappeared from 
the low-energy  region, and  energies of the
lowest-lying (deepest) states take  physically reasonable values.
It is to be emphasized that the binding energies of
so-obtained lowest-lying physical 
states are stable against further reduction 
of ${\widehat N}_{\rm max}$.

%\vskip 0.15cm
Table IV explicitly demonstrates {\it Step} ii).  We start with 
$\alpha_{\rm max}=4400$  
basis functions \{$\Phi_\alpha^{\rm (sym)}$\} 
whose parameters are given in Table II.
When the size of the new basis \{${\widehat \Phi}_N^{\rm (sym)}$\}
is reduced from $\alpha_{\rm max}$ to
${\widehat N}_{\rm max}=3250$ according to (2.37),
the solution of (2.35) has come to include no
unphysical state and give the binging energies 
$B_3^{(0)}=126.3999$ mK and $B_3^{(1)}=2.270606$ mK 
for the lowest two states.
By checking the stability of the
energy values against further decreasing  ${\widehat N}_{\rm max}$,
we verify the values of   
$B_3^{(0)}=126.40$ mK and $B_3^{(1)}=2.2706$ mK
in Table I.

%
%%%%%%%%%%%%%%%%%%%%%%%%%%%%%%%%%%%%%%%%%%%%%%%%%%%%%%%%%%%%
\section{$^4$H\lowercase{e} Tetramer}
%%%%%%%%%%%%%%%%%%%%%%%%%%%%%%%%%%%%%%%%%%%%%%%%%%%%%%%%%%%%

Calculation of the $^4$He tetramer using  realistic potentials
has been performed 
in Refs.~\cite{Carbonell,Lewerenz,Bressanini,Blume,Das}.
Although binding energy of the ground state  
obtained in the papers
agrees well with each other ($\sim\!558$ mK), 
that of the loosely bound excited state  differs  significantly 
from each other; namely, the binding energy  with respect to 
the trimer ground state (126.4 mK) is given as
1.1 mK by the Faddeev-Yakubovski (FY) 
equations method~\cite{Carbonell}, 
6.6 mK by  Monte Carlo methods combined
with the adiabatic hyperspherical approximation~\cite{Blume} 
and 52 mK recently by using a method of the correlated potential 
harmonic basis functions~\cite{Das}.  
Though the Faddeev result (1.1 mK) seems to the present authors 
the most accurate,  
the excited state was not solved 
as a bound-state problem 
in Ref.~\cite{Carbonell} % due to some difficulty 
but the result was extrapolated 
from the atom-trimer scattering phase shifts.

%\vskip 0.1cm
Thus the purpose of this section is
to perform, using the same LM2M2 potential as 
in Ref.~\cite{Carbonell}, accurate bound-state calculation
of the tetramer excited state, 
not only giving a precise binding energy 
but also describing the
short-range correlation and the asymptotic behavior of the
wave function properly.

The GEM has extensively 
been employed in bound-state calculations of
various four-body systems in nuclear and  
hypernuclear physics (cf. review papers \cite{Hiyama03,Hiyama09,
Hiyama10}).
Extension from three-body GEM calculations to four-body
ones in the presence of 
strong short-range repulsion is a familiar subject in nuclear physics.
For example, the study of three-nucleon 
bound states ($^3$H and $^3$He nuclei) 
in Ref.~\cite{Kameyama89} was extended to 
that of  four-nucleon ground state 
($^4$He nucleus, $J^\pi=0^+$)~\cite{Kamada01}
and the first excited, very diffuse 
state ($J^\pi=0^+$)~\cite{Hiyama03second}.
The  study of the three-$\alpha$-particle system 
($^{12}$C nucleus)~\cite{Hiyama03,Funaki,Kurokawa}
was extended to that of 
the four-$\alpha$-particle system ($^{16}$O nucleus)~\cite{Funaki}
with the strongly repulsive Pauli-blocking projection operator
on the $\alpha$-$\alpha$ motion.
Therefore, extension of the $^4$He trimer calculation 
to the tetramer one is straightforward
on account of those experiences.

%%%%%%%%%%%%%%%%%%%%%%%%%%%%%%%%%%%%%%%%%%%%%%%%%%%%%%%%%%%%
\subsection{Method}
%%%%%%%%%%%%%%%%%%%%%%%%%%%%%%%%%%%%%%%%%%%%%%%%%%%%%%%%%%%%

\begin{figure}[h]
\begin{center}
\epsfig{file=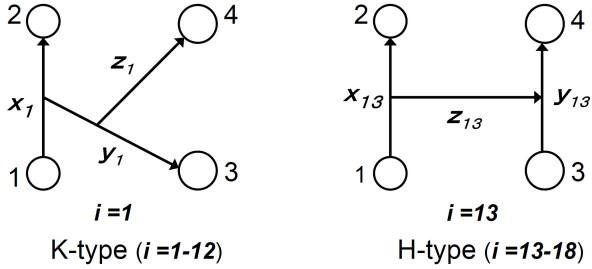,scale=0.18}
\end{center}
\caption{Jacobi coordinates, K-type and H-type, for the
$^4$He tetramer. Symmetrization of the four particles 
generates the sets $i=1, ..., 12$ (K-type) and 
$i=13, ...,  18$ (H-type).
}
\label{fig:4bodyjacobi}
\end{figure}

We take two types of Jacobi coordinate sets, K-type and H-type
(Fig.~\ref{fig:4bodyjacobi}).
Namely, for K-type, ${\bf x}_1={\bf r}_2-{\bf r}_1$,
${\bf y}_{1}={\bf r}_3 -\frac{1}{2}({\bf r}_1 + {\bf r}_2)$ and
${\bf z}_{1}={\bf r}_4 -\frac{1}{3}({\bf r}_1 +{\bf r}_2 + {\bf r}_3)$ 
and cyclically for $\{{\bf x}_i, {\bf y}_i, {\bf z}_i;\, i=2, ..., 12\}$ 
by the symmetrization between the four particles.
For H-type,
${\bf x}_{13}={\bf r}_2-{\bf r}_1$,
${\bf y}_{13}={\bf r}_4-{\bf r}_3$,
${\bf z}_{13}=\frac{1}{2}({\bf r}_3 + {\bf r}_4)
 -\frac{1}{2}({\bf r}_1 + {\bf r}_2)$ 
and cyclically for $\{{\bf x}_i, {\bf y}_i, {\bf z}_i;\, i=14, ..., 18\}$. 
An explicit illustration of the totally 18 sets 
of the rearrangement Jacobi coordinates
of four-body systems is seen in Fig.~18 of Ref.~\cite{Hiyama03}.

%%%%%%%%%%%%%%%%%%%%%%  Fig. 7   %%%%%%%%%%%%%
%%%%%%%%%%%%%%%%%%%%%%%%%%%%%%%%%%%%%%%%%%%

%\vskip 0.1cm
The total four-body  wave function $\Psi_4$ is
to be obtained by solving the Sch\"{o}dinger equation
\begin{equation}
( H - E ) \Psi_4 =0
\end{equation}
with the Hamiltonian 
\begin{equation}
H= -\frac{\hbar^2}{2\mu_x} \nabla^2_x
   -\frac{\hbar^2}{2\mu_y} \nabla^2_y
   -\frac{\hbar^2}{2\mu_z} \nabla^2_z
 + \sum_{1=i<j}^4 V(r_{ij}),
\label{eq:Hamil4}
\end{equation}
where $\mu_x=\frac{1}{2} m$, $\mu_y= \frac{2}{3} m$ and 
$\mu_z= \frac{3}{4} m$ on the K-type coordinates, and
$\mu_x=\mu_y=\frac{1}{2} m$ and $\mu_z=m$ on the H-type ones. 

$\Psi_4$ is
expanded in terms of the  symmetrized 
$L^2$-integrable K-type and H-type four-body
basis  functions:
%%%%%%%%%%%%%%%%%%%%%%%%%%%%%%%%%
\begin{equation}
 \Psi_4 =
\sum_{{\alpha_{\rm K}}=1}^{\alpha_{\rm K}^{\rm max}}
  A^{({\rm K})}_{\alpha_{\rm K}} 
\Phi^{({\rm sym;K})}_{\alpha_{\rm K}} +
\sum_{{\alpha_{\rm H}}=1}^{\alpha_{\rm H}^{\rm max}}
  A^{({\rm H})}_{\alpha_{\rm H}} \Phi^{({\rm sym;H})}_{\alpha_{\rm H}} ,
\end{equation}
%%%%%%%%%%%%%%%%%%%%%%%%%%%%%%%%%%
with
%%%%%%%%%%%%%%%%%%%%%%%%%%  
%\vskip -0.6cm
\begin{eqnarray}
 \Phi_{\alpha_{\rm K}}^{({\rm sym;K})} &=& \sum_{i=1}^{12} 
    \Phi^{({\rm K})}_{\alpha_{\rm K}}
( {\bf x}_i, {\bf y}_i, {\bf z}_i ), \\
 \Phi_{\alpha_{\rm H}}^{({\rm sym;H})} &=& \sum_{i=13}^{18} 
    \Phi^{({\rm H})}_{\alpha_{\rm H}}
( {\bf x}_i, {\bf y}_i, {\bf z}_i ), 
\end{eqnarray}
%%%%%%%%%%%%%%%%%%%%%%%%%%%%%%%
in which $\Phi( {\bf x}_i, {\bf y}_i, {\bf z}_i )$ is a function of
$i$-th set of Jacobi coordinates.
It is of importance that $ \Phi_{\alpha_{\rm K}}^{({\rm sym;K})} $
and $ \Phi_{\alpha_{\rm H}}^{({\rm sym;H})} $
are constructed on the full 18 sets of
Jacobi coordinates; this makes the function space of the basis 
quite wide.

%\vskip 0.1cm
The eigenenergies $E$ and amplitudes 
$A_{\alpha_{\rm K}}^{\rm (K)}  (A_{\alpha_{\rm H}}^{\rm (H)}) $ 
are determined
by the Rayleigh-Ritz variational principle:
\begin{eqnarray}
\langle \: \Phi_{\alpha_{\rm K}}^{({\rm sym;K})}
 \:| \:H - E \: |\: \Psi_4 \: \rangle =0, \\
\langle \: \Phi_{\alpha_{\rm H}}^{({\rm sym;H})}
 \:| \:H - E \: |\: \Psi_4 \: \rangle =0, 
\end{eqnarray}
where $\alpha_{\rm K}=1, ...,  \alpha_{\rm K}^{\rm max}$
and   $\alpha_{\rm H}=1, ...,  \alpha_{\rm H}^{\rm max}$.
This set of equations results in a generalized eigenvalue problem:
\begin{equation}
\sum _{c'={\rm K, H}} \;\;
\sum_{\alpha_{\rm c'}=1}^{\alpha_{\rm c'}^{\rm max}} 
\Big[ {\cal H}_{{\alpha_{\rm c}},{\alpha_{\rm c'}}}^{\rm (c,c')} -
     E\,{\cal N}_{{\alpha_{\rm c}},{\alpha_{\rm c'}}}^{\rm (c,c')}
 \Big] \:A_{\alpha_{\rm c'}}^{\rm (c')} = 0,
\end{equation}
where ${\rm c}= {\rm K , H}$ and 
$\alpha_{\rm c}=1, ...,  \alpha_{\rm c}^{\rm max}$.
The matrix elements are given by
\begin{eqnarray}
  {\cal H}_{{\alpha_{\rm c}},{\alpha_{\rm c'}}}^{\rm (c,c')} &=&  
\langle \: \Phi_{\alpha_{\rm c}}^{\rm (c)}  \:| \:H\: |\: 
\Phi_{\alpha_{\rm c'}}^{\rm (c')} \: \rangle ,\\
  {\cal N}_{{\alpha_{\rm c}},{\alpha_{\rm c'}}}^{\rm (c,c')} &=&  
\langle \: \Phi_{\alpha_{\rm c}}^{\rm (c)}  \:| \:\;1\:\; |\: 
\Phi_{\alpha_{\rm c'}}^{\rm (c')} \: \rangle .
\end{eqnarray}
%%%%%%%% 
Up to here is the
most general way of variational calculations
for bound states of identical spinless four particles.

We describe the basis function
$\Phi^{({\rm K})}_{\alpha_{\rm K}}(\Phi^{({\rm H})}_{\alpha_{\rm H}})$ 
in the form
%%%%%%%%%%%%%%%%%%%%%%% 
\begin{eqnarray}
 \Phi^{({\rm K})}_{\alpha_{\rm K}}( {\bf x}_i, {\bf y}_i, {\bf z}_i ) 
= \phi^{(^{\rm cos}_{\rm sin})}_{n_x l_x}(x_i) \,
  \phi_{n_y l_y}(y_i) \,
  \varphi_{n_z l_z}(z_i) \nonumber \\
\times \Big[ \big[Y_{l_x}({\widehat {\bf x}_i}) %\otimes
Y_{l_y}({\widehat {\bf y}_i}) \big]_\Lambda %\otimes
Y_{l_z}({\widehat {\bf z}_i}) 
  \Big]_{J M}, \nonumber \\
(i=1, ..., 12)
\end{eqnarray}
%%%
%\vskip -0.8cm
\begin{eqnarray}
 \Phi^{({\rm H})}_{\alpha_{\rm H}}( {\bf x}_i, {\bf y}_i, {\bf z}_i ) 
= \phi^{(^{\rm cos}_{\rm sin})}_{n_x l_x}(x_i) \,
  \psi_{n_y l_y}(y_i) \,
  \varphi_{n_z l_z}(z_i) \nonumber \\
\times \Big[ \big[Y_{l_x}({\widehat {\bf x}_i}) %\otimes
Y_{l_y}({\widehat {\bf y}_i}) \big]_\Lambda %\otimes
Y_{l_z}({\widehat {\bf z}_i}) 
  \Big]_{J M}, \nonumber \\
(i=13, ..., 18)
\end{eqnarray}
%%%%%%%%%%%%%%%%%%%%%%%%%%%%%%%%
%\vskip -0.2cm
where $\alpha_{\rm K}$ specifies a set 
\begin{eqnarray}
\!\!\!\alpha_{\rm K} \equiv \mbox{
 \{`cos' or `sin',} \, \omega,
n_x l_x, n_y l_y, n_z l_z , \Lambda,JM \},
\end{eqnarray}
which is commonly for the components  $i= 1, ..., 12$;
and similarly for $\alpha_{\rm H} $
commonly for $i=13, ..., 18$.

%\vskip 0.1cm
Since we consider the case of $J=0$ in this paper,
the totally symmetric four-body wave function
requires i)~$l_x={\rm even}$,  $l_y+l_z={\rm even}$ and
$\Lambda=l_z$ for the K-type basis and 
ii)~$l_x={\rm even}$,  $l_y={\rm even}$ and
$\Lambda=l_z={\rm even}$ for the H-type basis.

%\vskip 0.1cm
In (3.11) and (3.12), the radial functions are assumed,
as in Sec.II, to be 
\begin{eqnarray}
&&\phi^{(^{\rm cos}_{\rm sin})}_{n_x l_x}(x) =
x^{l_x}\:e^{- (x/x_{n_x})^2} \times
\big\{  \,^{{\rm cos}\, \omega (x/x_{n_x})^2}_{{\rm sin}\, 
\omega (x/x_{n_x})^2}  \;,\quad \\
&& \psi_{n_y l_y}(y)=
y^{l_y}\:e^{- (y/y_{n_y})^2}, \: \;  \\ 
&& \varphi_{n_z l_z}(z)=
z^{l_z}\:e^{-(z/z_{n_z})^2} \: \;   
\end{eqnarray}
with  geometric sequences of the Gaussian ranges:
\begin{eqnarray}
&& x_{n_x}= x_1\, a_x^{n_x-1}
\quad \:(n_x=1, ..., n_x^{\rm max})\;,  \\
&& y_{n_y}=y_1\, a_y^{n_y-1}
\quad \:(n_y=1, ..., n_y^{\rm max}),\; \\ 
&& z_{n_z}=z_1\, a_z^{n_z-1}
\quad \:(n_z=1, ..., n_z^{\rm max}).\;  
\end{eqnarray}
%
%\vskip 0.2cm
In (3.15),  the 'cos(sin)'-type function is not adopted
for $\psi_{n_y l_y}(y)$ 
of the H-type basis
though $y$ is the distance between two particles.
This is because  the 'cos(sin)'-type basis for the $x$-coordinate are
applied to all the pairs of H-type  by the symmetrization
of the four particles.

%\vskip 0.1cm
In the tetramer calculation, the total number,
$\alpha_{\rm max}=\alpha_{\rm K}^{\rm max}+\alpha_{\rm H}^{\rm max}$,  
of the symmetrized four-body basis functions (3.4) and (3.5) amounts 
to $\alpha_{\rm max}\sim$ 30000, ranging 
from very compact to very diffuse,
to obtain a well converged solution. Since the nonorthogonality
among those basis functions is
too large to solve directly the generalized eigenvalue problem (3.8),
we take the same two-step method as described in Sec.II.H in the
trimer calculation.   We finally solve the same type of 
standard eigenvalue problem
as (2.35) using the symmetric orthonormal four-body
basis functions, $\{ {\widehat\Phi}^{({\rm sym})}_N; 
N=1, ..., {\widehat N}_{\rm max} \}$,
in which the basis with too small 
$\nu_N$ have been omitted.

%%%%%%%%%%%%%%%%%%%%%%%%%%%%%%%%%%%%%%%%
\subsection{Binding energy}
%%%%%%%%%%%%%%%%%%%%%%%%%%%%%%%%%%%%%%%%%

In the calculation of the $^4$He-tetramer ground state
$\Psi_4^{(0)}$ and the excited state $\Psi_4^{(1)}$, 
the converged result was obtained by employing
the symmetric four-body basis functions of $\alpha_{\rm max}=29056$
with $l_{\rm max}=4$.
Table~V shows the convergence of the binding energies of
the tetramer ground $(B_4^{(0)})$ and excited $(B_4^{(1)})$ states
with respect to increasing $l_{\rm max}$.
Column 'K+H' is the result with both the K-type and H-type basis
functions in (3.3),
and column '(K)' is that with the K-type basis only.

%\vskip 0.1cm
Contribution from the K-type basis 
is dominant, but that from the H-type is sizable.
Without the latter the excited state does not become bound
($B_4^{(1)}<B_3^{(0)}=126.40$ mK) even for $l_{\rm max}=4$.
Since  both type bases are not
orthogonal to each other, the role of H-type one can be 
substituted in principle by the K-type one 
if a very large $l_{\rm max}$ is 
employed. But, this is not practical; use of both types of bases
is essentially important.

%\vskip 0.1cm
As seen in Table V, convegence of the binding 
energies with increasing $l_{\rm max}$
is more rapid in our calculation  than that
in the FY-equations calculation \cite{Carbonell}. 
The reason of this difference in the conversion
is the same as that mentioned in the trimer calculation (Sec.II.E).

%\vskip 0.1cm
%Nonlinear parameters of  the tetramer basis functions used for 
%the above calculation
%are listed in Appendix about the dominant part (23504 basis with
%$l_{\rm max}=2$).

%%%%%%%%%%%%%%%%   Table V  %%%%%%%%%%%%%
\begin{table}[t]
\caption{ Convergence of $^4$He tetramer calculations
with respect to the increasing maximum partial wave $(l_{\rm max})$.
$B_4^{(0)}$ and $B_4^{(1)}$ are
the binding energies of the
ground  and excited states, respectively.
Column 'K+H' is the results with both the K-type and H-type basis
functions  in (3.3) and column '(K)' is that with the K-type basis only.
The excited state is unbound 
($B_4^{(1)}<B_3^{(0)}=126.40$ mK) in the cases denoted 
by the symbol ' $-$ '.
For comparison, $B_4^{(0)}$  obtained by the FY-equations 
calculation~\cite{Carbonell}
is listed in the last column.
The LM2M2 potential is employed.
}
\label{table:1}
\begin{center}
\begin{tabular}{cccccccc} 
\hline \hline
\noalign{\vskip 0.1 true cm} 
tetramer  & \multicolumn{5}{c} {present}  &  $\;$
  &  \multicolumn{1}{c}  {Ref.\cite{Carbonell}}   \\
\noalign{\vskip -0.2 true cm} 
 &  \multispan5 {\hrulefill} &  & \multispan1 {\hrulefill}
   \cr
\noalign{\vskip 0.0 true cm} 
& \multicolumn{2}{c}{$B_4^{(0)}$ (mK)} &  
& \multicolumn{2}{c}{$B_4^{(1)}$ (mK)} & & 
$B_4^{(0)}$ (mK)   \\
\noalign{\vskip -0.2 true cm} 
 & \multispan2 {\hrulefill} & & \multispan2 {\hrulefill} & & \\
\noalign{\vskip 0.0 true cm} 
 $l_{\rm max}$   
 & K+H & ( K ) & & K+H & ( K ) & & \\
\noalign{\vskip 0.1 true cm} 
\hline 
\noalign{\vskip 0.2 true cm} 
  0   &  500.71 & $\,$ (185.96) &$\quad$  & $-$ & $\,$ ( $-$ ) &  & 348.8 \\
  2   &  558.29 & $\,$ (508.62) &  &   127.24 & $\,$ ( $-$ )  &  &  505.9    \\
  4   &  558.98 & $\,$ (532.56) &  &   127.33 & $\,$ ( $-$ )  &  &  548.6    \\
  6   &    &  & & &       &  &  556.0   \\
 8   &     &  & & &       &   &  557.7  \\
%\vspace{-5 mm} \\
\noalign{\vskip 0.1 true cm} 
\hline
\hline
\end{tabular}
\label{table:B(4)}
\end{center}
\end{table}
%%%%%%%%%%%%%%%%%%%%%%%%%%%%%%%%%%%%%%

%%%%%%%%%%%%%%%%%%%%%%%%%%%%%%  Table VI  %%%%%%%%%%%
\begin{table}  [h]
\caption{ Nonlinear parameters of the four-body 
Gaussian basis functions, (3.11)--(3.19), 
used for the $^4$He tetramer ground  and excited
states with $J=0 \, (\Lambda=l_z)$
in the case of $l_{\rm max}=2$ with 23504 basis 
functions (cf. Table V). 
In the leftmost column, K(H) stands for
the K-type (H-type) basis.
We take $\omega=1.0$ in (3.14). 
}
\begin{center}
\begin{tabular}{ccccccccccccccc}
\hline
\hline
\noalign{\vskip 0.2 true cm} 
   \multicolumn{1}{c}{}  & \multicolumn{4}{c} 
{ $\phi^{({\rm cos})}_{n_x l_x}, \phi^{({\rm sin})}_{n_x l_x}$ } 
& &  \multicolumn{4}{c} { $\psi_{n_y l_y}$}   
& &  \multicolumn{4}{c} { $\varphi_{n_z l_z}$}   \\
%\noalign{\vskip 0.1 true cm}
   &
   \multispan4 {\hrulefill} &  &   \multispan4 {\hrulefill} &
    &   \multispan4 {\hrulefill}  \\
\noalign{\vskip 0.1 true cm} 
 & $l_x $ & $n_x^{\mbox {\tiny {\rm max}}}$ &$x_1$ & $x_{n_x^{\rm max}}$
& & $l_y$ & $n_y^{\mbox {\tiny {\rm max}}}$ &$y_1$ & $y_{n_y^{\rm max}}$
& & $l_z$ & $n_z^{\mbox {\tiny {\rm max}}}$ &$z_1$ & $z_{n_z^{\rm max}}$
\\
%\vspace{-3 mm} \\
%\noalign{\vskip 0.1 true cm} 
  &      &              &      [\AA] &             [\AA]  
& \qquad &       &      &       [\AA] &            [\AA] 
& \qquad &       &      &       [\AA] &            [\AA] 
\\
\noalign{\vskip 0.1 true cm} 
\hline
\vspace{-3 mm} \\
K & 0   &  14  &  0.2 &   20.0 &  & 0  &  15  &  0.8 & 50.0  &
  & 0  &  20  &  0.8 &   400.0   \\
%2
K & 2   &  12  &  0.4 &   20.0 &  & 2  &  14  &  0.8 & 40.0  &
  & 0  &  16  &  0.8 &   300.0   \\
%3
K & 0   &   8  &  0.3 &    6.0 &  & 2  &   8  &  0.8 &  6.0  &
  & 2  &   8  &  1.0 &     6.0   \\
%4
K & 2   &   6  &  0.3 &    6.0 &  & 0  &   8  &  0.8 &  6.0  &
  & 2  &   8  &  1.0 &     6.0   \\
%5
K & 0   &   8  &  0.3 &    6.0 &  & 1  &   8  &  0.8 &  6.0  &
  & 1  &   8  &  1.0 &     6.0   \\
%6
K & 2   &   6  &  0.4 &    6.0 &   & 1  &   6  &  0.8 &  6.0  &
  & 1  &   8  &  1.0 &     6.0   \\
%7
K & 2   &   6  &  0.5 &    6.0 &   & 2  &   8  &  0.8 &  6.0  &
  & 2  &   8  &  1.0 &     6.0   \\
%8
H & 0   &  12  &  0.3 &   20.0 &   & 0  &  12  &  0.4 & 16.0  &
  & 0  &  14  &  0.8 &   25.0   \\
%9
H & 2   &   6  &  0.6 &    6.0 &   & 2  &   6  &  0.8 &  6.0  &
  & 0  &   8  &  1.0 &     6.0  \\
%10
H & 0   &   6  &  0.3 &    6.0 &   & 2  &   6  &  0.8 &  6.0  &
  & 2  &   8  &  1.0 &     6.0  \\
%11
H & 2   &   6  &  0.6 &    6.0 &   & 0  &   6  &  0.8 &  6.0  &
 & 2  &   8  &  1.0 &     6.0  \\
%12
H & 2   &   6  &  0.6 &    6.0 &   & 2  &   6  &  0.8 &  6.0  &
  & 2  &   8  &  1.0 &     6.0  \\
\vspace{-3 mm} \\
\hline
\hline
\end{tabular}
\end{center}
%}
\end{table}
%%%%%%%%%%%%%%%%%%%%%%%%%%%%%%%%%%%%

%%%%%%%%%%%%%%%%%%%%   Table VII   %%%%%%%%%
\begin{table}[h]
\caption{ Stability of the calculated 
binding energies of the $^4$He tetramer
ground and excited states with respect to 
the number (${\widehat N}_{\rm max}$) 
of the symmetrized orthonormal four-body basis functions 
corresponding  to 
\{${\widehat \Phi}_N^{\rm (sym)}$\} in Eq.(2.33)
in the case of $l_{\rm max}=4$ with the LM2M2  potential.
This concludes $B_4^{(0)}=558.98$ mK and $B_4^{(1)}=127.33$ K. 
}
\label{table:result}
\begin{center}
\begin{tabular}{cccc}
\hline
\hline
\vspace{-3 mm} \\
  $ {\widehat N}_{\rm max}$ & $(\Delta {\widehat N}_{\rm max}) $
 &   $B_4^{(0)}$ (mK)  & $B_4^{(1)}$ (mK)  \\
\vspace{-3 mm} \\
\hline
\vspace{-3 mm} \\
 24800 &  $  -$  &  $\quad 558.983 \quad$ & $ 127.326 $\quad   \\
 24790 & $( -10)$&  $ 558.981 $ & $ 127.326 $   \\
 24780 & $( -20)$&  $ 558.980 $ & $ 127.326 $   \\
 24750 & $( -50)$&  $ 558.977 $ & $ 127.326 $   \\
 24700 & $( -100)$&  $558.975 $ & $ 127.325 $   \\
 24300 & $( -500)$&  $558.954 $ & $ 127.323 $   \\
 23800 & $(-1000)$&  $558.924 $ & $ 127.320 $   \\
%22800 & $(-2000)$&  $558.853 $ & $ 127.312 $   \\
\vspace{-3 mm} \\
\hline
\hline
\end{tabular}
\end{center}
\end{table}
%%%%%%%%%%%%%%%%%%%%%%%%%%%%%%%%%%%%

%\vskip 0.1cm
In Table  VI we list the nonlinear parameters 
of the four-body Gaussian basis functions, (3.11)--(3.19), 
in the case of $l_{\rm max}=2$ with 23504 basis 
functions (cf. Table V) to avoid too long listing for $l_{\rm max}=4$.
The range parameters are given in  round numbers
but further optimization of them do not improve the 
binding energies 
as long as we calculate them with five
significant figures.

When solving the generalized eigenvalue problem (3.8), we take
the same two-step method as mentioned in Sec.II.H.
Table VII shows stability of the binding energes of the
ground and excited states $(B_4^{(v)}, v=0,1)$
against the decreasing number  ${\widehat N}_{\rm max}$ of the symmetrized
orthonormal four-body basis functions
corresponding  to 
\{${\widehat \Phi}_N^{\rm (sym)}$\} in Eq.(2.33).

%\vskip 0.1cm

Calculated binding energies and some of mean values of the tetramer 
ground and excited states 
are summerized in Table VIII(a) in comparison with those obtained 
by Lazauskas and Carbonell \cite{Carbonell} with
the FY-equations method in which
the excited state was not  obtained by 
a direct bound-state calculation 
but the binding energy (127.5 mK) was extrapolated
from the  atom-trimer scattering calculations.
Our result of $B_4^{(1)}=127.33 $ mK, which is very 
closed to $B_4^{(1)}$
in Ref.\cite{Carbonell}, confirms
the existence of the very shallow
bound excited state ($J=0$) of the $^4$He tetramer.
The tetramer excited state is located only by 0.93 mK below the
trimer ground state ($126.40$ mK). This is analogous to
that the trimer excited state lies by 0.967 mK below 
the dimer; a reason was explained in Sec.II.G by taking 
the dimerlike-pair model.

%\vskip 0.15cm
Table VIII(b) lists $B_4^{(0)}$ and $B_4^{(1)}$ 
obtained in other literature papers
by the Monte Carlo 
methods~\cite{Lewerenz,Blume,Bressanini}
and by using the correlated potential harmonic 
basis~\cite{Das}. All $B_4^{(0)}$ values
agree well with the results  by the present 
and FY-equations calculations, but $B_4^{(1)}$ by 
Refs.~\cite{Blume,Das} deviate significantly
from our and FY-equations results.

%\vskip 0.15cm
Use of $\frac{\hbar^2}{m}=12.11928$ K\AA$^2\,$~\cite{newmass}
results in  $B_4^{(0)}=559.22$  mK 
and $B_4^{(1)}=127.42$  mK.
The same perturbative treatment for the small difference of 
$\frac{\hbar^2}{m}$ as used in Sec.II.E gives
$B_4^{(0)}=559.23$ mK and
$B_4^{(1)}=127.42$ mK.
The calculations below in Sec.III.C take 
$\frac{\hbar^2}{m}=12.12$ K\AA$^2$.

%%%%%%%%%%%%%%%%%%%%%%  Table VIII    %%%%%%%
%
\begin{table} [t]  
\begin{center}
\caption{ (a) Mean values of the $^4$He tetramer ground 
and excited states calculated by the present work and 
the FY-equations method~\cite{Carbonell} with the use of 
the LM2M2 potential.
$B_4$ is  the binding energy, $r_{ij}$ stands for interparticle
distance and $r_{i{\rm G}}$ is the distance of a particle from the
center-of-mass of the tetramer. (b) The other literature 
work~\cite{Lewerenz,Bressanini,Blume,Das} on the binding 
energies (see text). 
The values originally given in units of cm$^{-1}$  are
transformed in units of mK in the parentheses.
}
\begin{minipage}[h]{8.0cm}
\begin{tabular}{crrrrr} 
 (a) \\
\hline 
\hline
\noalign{\vskip 0.1 true cm} 
tetramer   & \multicolumn{2}{c} {ground state} 
 & & \multicolumn{2}{c} {excited state}  \\
\noalign{\vskip -0.2 true cm} 
 &  \multispan2 {\hrulefill} & & \multispan2 {\hrulefill} \\
%\hline  
%\noalign{\vskip 0.1 true cm} 
 & \quad present  & \quad Ref.\cite{Carbonell}
  & &\quad present 
& \quad Ref.\cite{Carbonell}  \\
\noalign{\vskip 0.1 true cm} 
\hline 
\noalign{\vskip 0.1 true cm} 
  $B_4$  (mK) 
&   558.98   &  557.7  & $\quad$ &  127.33  &  127.5  \\
 $\langle T \rangle $  (mK)   
&  4282.2    &  4107  & &  1639.2      &  \\
 $\langle V \rangle $  (mK)
&  $-4841.2$   &  $-4665$ &  & $-1766.5$          &  \\
 $ \sqrt{ \langle r^2_{ij} \rangle  }$ (\AA)
&  $ 8.43$   &  8.40   & & 54.5   &  34.4 \\
 $  \langle r_{ij} \rangle  $ (\AA)
&  $7.70$   &        &  & 35.8   &  \\
 $  \langle r_{ij}^{-1} \rangle  $ (\AA$^{-1})$
&  $ 0.155 $   &        &  &  0.0792  &  \\
 $  \langle r_{ij}^{-2} \rangle  $ (\AA$^{-2})$
&  $ 0.0285 $   &         &  & 0.0117    &  \\
 $ \sqrt{ \langle r_{i{\rm G}}^{2} \rangle } $ (\AA)
&  5.16   &     & &  33.3 &      \\
\noalign{\vskip 0.1 true cm} 
  $C_4^{(v)}$(\AA$^{-\frac{1}{2}})$ 
&  2.1   &     & &  0.10  &      \\
\noalign{\vskip 0.1 true cm} 
\hline
%\noalign{\vskip 0.3 true cm} 
%% $^{*)}$See text.
%\noalign{\vskip 0.1 true cm} 
\label{table:5}
\end{tabular}
\end{minipage}
\begin{minipage}{8.0cm}
\begin{tabular}{ccccc} 
 (b) \\
\hline 
\hline 
\noalign{\vskip 0.1 true cm} 
tetramer & Ref.~\cite{Lewerenz}  & Ref.~\cite{Bressanini} 
& Ref.~\cite{Blume} & Ref.~\cite{Das} \\
\noalign{\vskip 0.1 true cm} 
\hline 
\noalign{\vskip 0.1 true cm} 
  $B_4^{(0)}$  (cm$^{-1}$) 
& $\,$ 0.388(1) $\,$ & $\,$ 0.3886(1) $\,$ 
& $\,$ 0.387(1) $\,$ & $\,$ 0.388 $\,$   \\
\qquad $$ (mK) 
&   (558)   &  (559.1)  & (557)  & (558)  \\
\noalign{\vskip 0.2 true cm} 
  $B_4^{(1)}$  (cm$^{-1}$) 
&      &    &  0.0922  &  0.124  \\
\qquad $$ (mK) 
&      &    & (133)  & (178)  \\
\noalign{\vskip 0.1 true cm} 
\hline
\hline
\noalign{\vskip -0.3 true cm} 
\end{tabular}
\end{minipage}
\end{center}
\end{table}
%%%%%%%%%%%%%%%%%%%%%%%%%%%%%%%%%%%%%%

%%%%%%%%%%%%%%%%%%%%%%%%%%%%%%%%%%%%%%%%
\subsection{Short-range correlation and asymptotic behavior}
%%%%%%%%%%%%%%%%%%%%%%%%%%%%%%%%%%%%%%%

Definition of the  pair correlation  function
(2.23) for the trimer is extended to  the tetramer states
$\Psi_4^{(v)} (v=0,1$):
\begin{equation}
P_4^{(v)}(x_1)\,Y_{00}({\widehat {\bf x}}_1) = 
\langle   \,\Psi_4^{(v)}\,| \,\Psi_4^{(v)}\,  
\rangle_{{\bf y}_1, {\bf z}_1} ,
\end{equation}
where 
$\langle \quad \rangle_{{\bf y}_1, {\bf z}_1}$ means
the integration over ${\bf y}_1$ and ${\bf z}_1$. 

%\vskip 0.1cm
The overlap function between  a tetramer state $\Psi_4^{(v)}$ 
and a trimer one $\Psi_3^{(v_3)} (v_3=0,1$)
is defined as  a function of the atom-trimer distance $z$ as an extension
from (2.24):
\begin{equation}
{\cal O}_4^{(v_3, v)}(z_1)\,Y_{00}({\widehat {\bf z}}_1) =  
\langle \,  \Psi_3^{(v_3)} \, | \,
\Psi_4^{(v)} \,\rangle_{{\bf x}_1, {\bf y}_1} .
\end{equation}

%%%%%%%%%%%%%%%%%%%%   Fig. 8  %%%%%%%%%%%%%%%
\begin{figure}[t]
\begin{center}
\epsfig{file=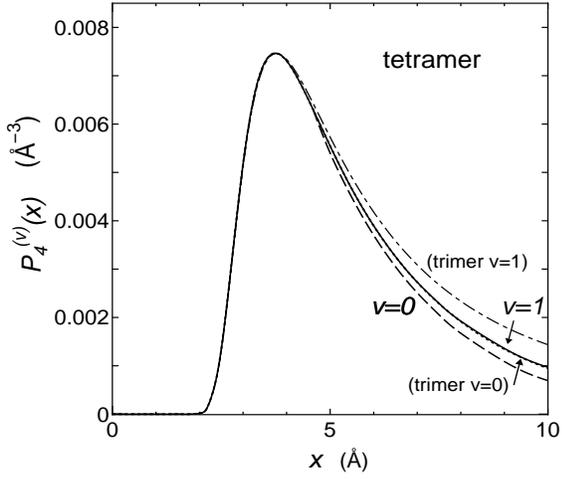,width=7.5cm,height=6.4cm}
\end{center}
\caption{
Short-range structure of the pair correlation function
$P_4^{(v)}(x)$ of the $^4$He tetramer calculated by (3.20).
The dashed line stands for the tetramer ground $(v=0)$ state
and the solid line for  the excited $(v=1)$ state.  
For the sake of comparison, additionally  shown are  
the dotted line for the trimer ground  state
and the dash-dotted line for the trimer excited state. 
The solid,  dotted and dash-dotted lines have been multiplied by  
factors 2.76, 1.36 and 19.8, respectively;
the same shape of the short-range correlation ($x \lesssim 4 $ \AA $\,$)
appears in all the states 
(cf. Fig.~\ref{fig:trimer-den-short} for dimer).
}
\label{fig:tetra-den-short}
\end{figure}
%%%%%%%%%%%%%%%%%%%%%%%%%%%%%%%%%%%%%%%%%%% 

%\vskip 0.1cm
All the  integrals in (3.20) and (3.21) give the analytical expression
owing to the use of Gaussian basis functions;
$\Psi_4^{(v)}$  is to be transformed to a function of
$({\bf x}_1, {\bf y}_1, {\bf z}_1)$, and 
$\Psi_3^{(v_3)}$ to that of $({\bf x}_1, {\bf y}_1)$.

%\vskip 0.1cm
In Fig.~\ref{fig:tetra-den-short}, 
we illustrate the short-range structure of
the pair correlation functions
$P_4^{(v)}(x)$ of the tetramer ground $(v=0)$ and excited $(v=1)$ states
together with those of the trimer states. 
It is to be emphasized that the same shape of short-range correlation
($x \lap 4 \,$ \AA) appears in all the states  (cf. Fig.~3 for the dimer)
without introducing any  pair correlation function.

%\vskip 0.1cm
The pair correlation functions $P_4^{(v)}(x)$ of the tetramer $(v=0,1)$
take   very small values 
in the strongly-repulsive potential 
region ($x \lap 1.5$ \AA) in Fig.~\ref{fig:tetra-den-short};
relative ratio of the values 
to the peak value is $\sim \!\! 10^{-6}$. This ratio is to be compared
with  $\sim \!\! 10^{-2}$ in the case of
the four-nucleon bound state
($^4$He nucleus) calculated with a realistic nucleon-nucleon
interaction with a strong short-range repulsion
(the ratio is  seen in  Fig.~1 
of Ref.~\cite{Kamada01} for the calculated 
pair correlation function of the $^4$He nucleus).$\;$ 
We understand that
so strong is the repulsive core of the atom-atom potential.

\begin{figure}[t]
\begin{center}
\epsfig{file=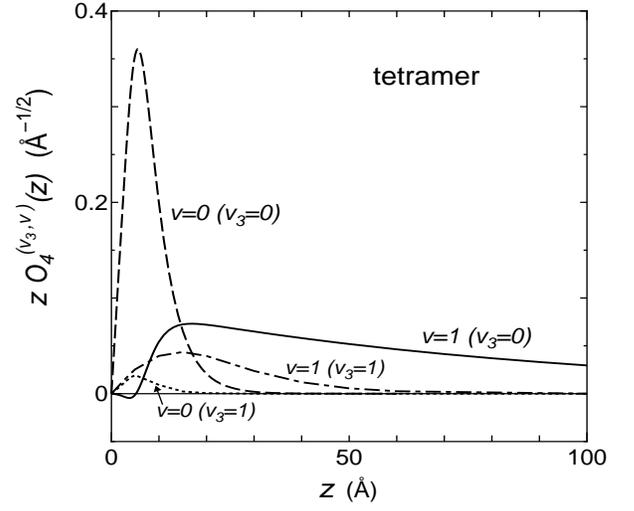,width=8.0cm,height=6.8cm}
\end{center} 
\caption{Overlap function ${\cal O}_4^{(v_3, v)}(z)$ in (3.21), 
multiplied by $z$,  between
the trimer state ($v_3=0, 1$) and
the tetramer state ($v=0, 1$) 
as a function of the atom-trimer distance $z$.
}
\label{fig:tetra-red-short}
\end{figure}

%\vskip 0.1cm
In Fig.~\ref{fig:tetra-red-short}, 
we plot the overlap function
${\cal O}_4^{(v_3, v)}(z)$, multiplied by $z$, 
 between the tetramer states ($v=0, 1$)
and the trimer states ($v_3=0, 1$) in the region $z \leq 100$ \AA.  
The two lines of $v=0, 1 (v_3=0)$
are to be compared with
the result in Fig.~4 of Ref.\cite{Carbonell} of 
the FY-equations calculation;
the latter result represents 
the K-type FY components as a function of atom-trimer distance $z$.
The two kinds of the  results are resemble to each other
though they do not stand for the same quantity. 
As for the excited state, the FY component is derived approximately
by modifying the  K-type FY amplitude of the zero-energy 
scattering~\cite{Carbonell}; the resulting amplitude 
is slightly more enhanced in the inner region than
our overlap function of $v=1 (v_3=0)$.
This is reflected in the r.m.s distance 
 $ \sqrt{ \langle r^2_{ij} \rangle  }$ in Table VIII(a).
  In the plot of 
the overlap functions between the trimer {\it excited} state $(v_3=1)$
and the tetramer states,
it is reasonably seen that ${\cal O}_4^{(v_3=1, v)}(z)$ 
is much smaller   than ${\cal O}_4^{(v_3=0, v)}(z)$ $\: (v=0,1)$
and  decreases more rapidly along $z$.

%%%%%%%%%%%%%%%%%%%%%%%%%%%  Fig. 9  %%%%%%%%%%%5%%%%%%%%
%%%%%%%%%%%%%%%%%%%%%%%%%%%%%%%%%%%%%%%%%%%

\begin{figure}[t]
\begin{center}
\epsfig{file=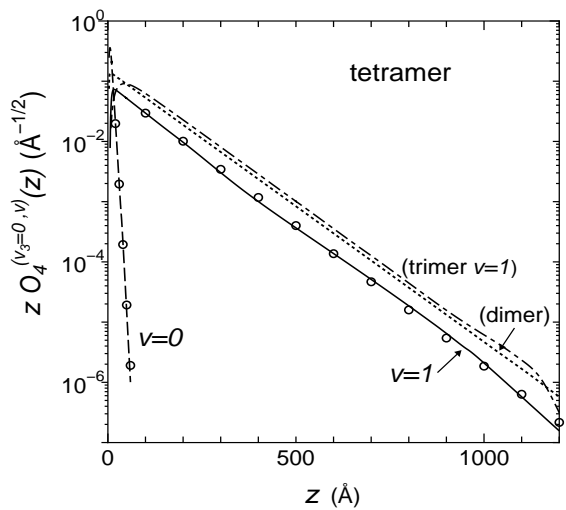,width=7.5cm,height=6.8cm}
\end{center}
\caption{
Asymptotic behavior of the overlap function 
${\cal O}_4^{(v_3=0, v)}(z)$, multiplied by $z$,
between the trimer ground state $(v_3=0)$
and the tetramer states $(v=0,1)$.
Open circles represent
the fit of the asymptotic function (3.23) to ${\cal O}_4^{(v_3=0, v)}(z)$
using the asymptotic normalization coefficient $C_4^{(v)}$.
}
\label{fig:tetra-red-long}
\end{figure}

%\vskip 0.1cm
In Fig.~\ref{fig:tetra-red-long}, 
$z\, {\cal O}_4^{(v_3=0, v)}(z) \,(v=0,1)$
are illustrated in the asymptotic region.
They should  satisfy
\begin{eqnarray}
z \,{\cal O}_4^{(v_3=0,v)}(z) 
\stackrel{z \to \infty}{-\!\!\!\longrightarrow}
C_4^{(v)}{\rm exp}(-\kappa_4^{(v)} z) 
\end{eqnarray}
with
$\kappa_4^{(v)}=\sqrt{2\mu_z (B_4^{(v)}-B_3^{(0)})}/\hbar$
($\kappa_4^{(0)}\!=\! 0.231 $ \AA$^{-1}$ and
$\kappa_4^{(1)}= 0.0107$ \AA$^{-1}$). 
The dashed line ($v\!=\!0$) and solid line ($v\!=\!1$) reproduces the
asymptotic functions (3.22) with the ANC 
$C_4^{(0)}\!=2.1\!$ \AA$^{-\frac{1}{2}}$ and
$C_4^{(1)}\!=0.10\!$ \AA$^{-\frac{1}{2}}$ (see the open circles), 
respectively, up to $y\! \sim\! 1000$ \AA.
%though the agreement in the excited state is
%slightly worse than that in the trimer.

%\vskip 0.1cm
Asymptotically the tetramer is dissociated 
into the trimer ground state and a distant atom in the symmetric way
between the four atoms (with negligible amount of the trimer
excited state). Therefore, the tetramer wave function 
$\Psi_4^{(v)} (v=0, 1)$ is represented asymptotically as
\begin{eqnarray}
\Psi_4^{(v)} 
\longrightarrow C_4^{(v)} \sum_{n=1}^{4}  \Psi_{3,n}^{(0)}\,
\frac{e^{-\kappa_4^{(v)} z_n}}{z_n}
  Y_{00}({\widehat {\bf z}_n}) ,
\end{eqnarray}
where the summation over $n$ symmetrizes the four atoms;
namely, the $n$th atom is isolated at ${\bf z}_n$ from
the trimer $\Psi_{3,n}^{(0)}$ in which the 
$n$th atom is absent and the other three atoms are symmetrized.

As for the  tetramer's ANC, we can make the same comments
as those for the trimer's ANC below Eq.(2.26).
%
%%%%%%%%%%%%%%%%%%%%%%%%%%  Fig. 10  %%%%%%%%%%%%%%%%
%%%%%%%%%%%%%%%%%%%%%%%%%%%%%%%%%%%%%%%%%%%

%%%%%%%%%%%%%%%%%%%%%%%%%%%%%%%%%%%%%%%%%%%%%%%%%%%%%%%%%%%%%%
\subsection{'Dimerlike-pair' model in asymptotic region}
%%%%%%%%%%%%%%%%%%%%%%%%%%%%%%%%%%%%%%%%%%%%%%%%%%%%%%%%%%%%%%

In Fig.~\ref{fig:tetra-red-long},
we note that the exponentially-decaying slope of the solid line
($\kappa_4^{(1)}$) is very close to that in the
trimer excited state ($\kappa_3^{(1)}$)
and that in the dimer ($\kappa_2$).
This gives a support to the dimerlike-pair model
for the tetramer excited state in the asymptotic region 
(Fig.~\ref{fig:dimer-model}b).

\begin{figure}[t]
\begin{center}
\epsfig{file=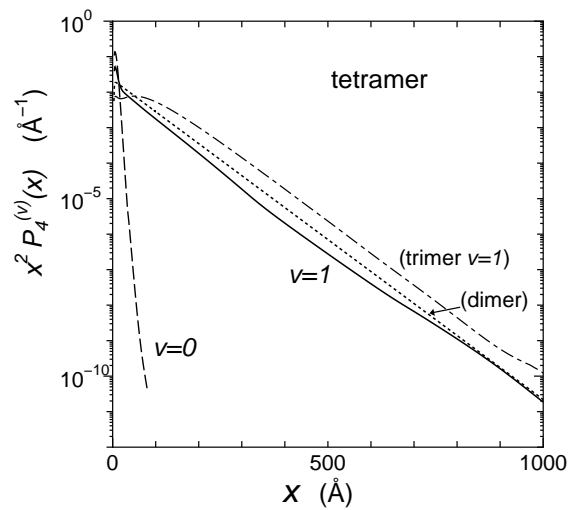,width=7.5cm,height=6.8cm}
\end{center}
\caption{
Asymptotic behavior of the pair correlation function
$P_4^{(v)}(x)$, multiplied by $x^2$, of the $^4$He tetramer
calculated by (3.20). 
The dashed line stands for the tetramer ground ($v=0$) state
and the solid line for the excited $(v=1)$ state.
The dotted line for the trimer excited state
and the dotted line for the dimer are added for the sake of
comparison. 
}
\label{fig:tetra-den-long}
\end{figure}

%\vskip 0.1cm
We can therefore predict that, in the asymptotic region,
 the pair correlation function
$x^2 P_4^{(1)}(x)$ should be proportional to 
${\rm exp}(-2\kappa_4^{(1)}x)$.
This is seen in Fig.~\ref{fig:tetra-den-long} 
though the solid line ($v=1$) is not so excellently straight 
as in the trimer excited state (dot-dashed line)
due to the complexity
of the four-body calculation.
 
%%%%%%%%%%%%%%%%    Fig. 11  %%%%%%%%%%%%%%%%%%%
%%%%%%%%%%%%%%%%%%%%%%%%%%%%%%%%%%%%%%%%%%%

%\vskip 0.1cm
As mentioned in Sec.II.G, the dimerlike-pair model predicted
$B_4^{(1)} \simeq 127.27$ mK. In the present  calculation,
we obtained  $B_4^{(1)}=127.33$ mK. The prediction is very good
as long as the LM2M2 potential is employed.

%\vskip 0.1cm
Generally, in the case of $^4$He$_N$,
the excited-state binding energy calculated using realistic potentials
%%ground state, say $\Delta B_N^{(1)}(=B_N^{(1)}-B_{N-1}^{(0)})$,
may be expected as  
\begin{equation}
B_N^{(1)} \simeq B_{N-1}^{(0)} +  \frac{N}{2(N-1)} B_2 ,
\end{equation}
where the factor multiplied to $B_2$ comes from the ratio of
the reduced mass of the dimerlike pair ($\frac{1}{2}m$) 
to that of the $^4$He$\,$-$^4$He$_{N-1}$ system ($\frac{N-1}{N}m$).
But, as discussed in Sec.II.G, the factor might
depend on the realistic potentials with a small deviation
(roughly $\pm 0.3$).

% We note that this {\it intuitive} dimerlike-pair model
% was generated by analyzing the {\it quantitative} results of
% the precise three- and four-body calculations.

%%%%%%%%%%%%%%%%%%%%%%%%%%%%%%%%%%%%%
\section{Summary}
%%%%%%%%%%%%%%%%%%%%%%%%%%%%%%%%%%%%%

We have calculated
the ground and excited states of $^4$He trimer and tetramer
using the LM2M2 potential which has a strong short-range repulsive 
potential and is one of the most widely used $^4$He-$^4$He interactions.
We employed the Gaussian expansion method (GEM) for {\it ab initio} 
variational calculations of few-body systems.
The symmetrized three-(four-)body Gaussian basis functions, 
ranging from very compact  to very diffuse,
are constructed on the full sets of possible Jacobi coordinates.
Therefore, the basis set, spanning a wide function space, is
suitable for describing both the short-range correlation
(without assuming any pair correlation function)
and the long-range asymptotic behavior of the trimer (tetramer) 
wave function as well as suitable 
for obtaining accurate binding energies. 
The main conclusions are summarized as follows:

%\vskip 0.1cm
i) Calculated binding energies of the trimer ground and 
excited states, $B_3^{(0)}$ and $B_3^{(1)}$, respectively,
agree excellently  with the literature (Table I); we have 
$B_3^{(0)}=126.40$ mK and $B_3^{(1)}=2.2706$ mK.  

%\vskip 0.1cm
ii) As for the binding energies of the tetramer ground 
and excited states,
we obtained $B_4^{(0)}=558.98$ mK and $B_4^{(1)}=127.33$ mK
(situated only 0.93 mK below the atom-trimer threshold).
The former is in good agreement with the literature calculations,
while the latter supports the result of 127.5 mK by 
Ref.~\cite{Carbonell} differently from the other 
literature results (Table VIII).

%\vskip 0.1cm
iii) We found that the strong
short-range correlation ($r_{ij} \lap 4$ \AA) seen
in the dimer appears also in the ground and excited states
of the trimer and tetramer precisely in 
the same shape (Figs.~3 and 8).
This gives a foundation to
an {\it a priori} assumption that a pair 
correlation function 
to simulate the short-range part of
the dimer wave function 
is incorporated in the three-(four-)body  
wave function from the beginning.

iv) Illustrating the overlap function  between
the trimer excited state and the dimer (${\cal O}_3^{(v=1)}$) 
and that between the tetramer excited state and the trimer
ground state (${\cal O}_4^{(v_3=0, v=1)}$), 
we found that those overlap functions are
almost proportional to the dimer wave function
in the asymptotic region up to $\sim 1000$ \AA $\,$ (Figs.~4 and 10).
Also it was found that the pair correlation functions
of trimer and tetramer excited states 
($P_3^{(v=1)}$ and $P_4^{(v=1)}$, respectively)
are almost proportional to
the squared dimer wave function in the asymptotic region 
(Figs.~6 and 11).
%% $\kappa_3^{(1)} \simeq \kappa_2$ and $\kappa_4^{(1)} \simeq \kappa_2$.
%%(with a binding wave number $\kappa_3^{(1)}$)
%% (with $\kappa_4^{(1)}$)
We then came to propose 
a 'dimerlike pair' model (Fig.~5)
that predicts the excited-state binding energy of $^4$He$_N$
($N \ge 3$) using Eq.~(3.24).
It will be of interest to examine this model
in the case of $N \ge 5$.
A five-body calculation of the pentamer,  $^4$He$_5$, is in progress.

%\vskip 0.1cm
v) We calculated the 
asymptotic normalization coefficient (ANC) of the 
tail function
of the \mbox{dimer-atom} \mbox{(trimer-atom)} relative motion
in trimer (tetramer).
This result may be available in peripheral reactions
(insensitive to the interior of the system) 
including the $^4$He trimers and
tetramers, in which the reaction cross section will be
proportional to the squared ANC. 
The ANC is a quantity  
to convey the interior structural information 
to the asymptotic behavior.
%reflect the internal structural
%information of trimer (tetramer).
Therefore, attention to this quantity might be helpful 
when one tries to reproduce the non-universal variation of the 
$^4$He trimer (tetramer) states 
by means of parametrizing \mbox{effective} models beyond
Efimov's universal theory.

\section*{Acknowledgement}
The numerical calculations were performed on \mbox{HITACHI SR16000}
at KEK and YIFP.

%%%%%%%%%%%%%%%%%%%%%%%%%%%%%%%%%%%%%%%%%%%%%%%%%%%%%%%%%%%%%%%%%%%%%%
%%%%%%%%%%%%%%%%%%%%%%%%%%%%%%%%%%%%%%%%%%%%%%%%%%%%%%%%%%%%
%%%%%%%%%%%%%%%%%%%%% Reference  %%%%%%%%%%%%%%%%%%%%%%%%%%%
%%%%%%%%%%%%%%%%%%%%%%%%%%%%%%%%%%%%%%%%%%%%%%%%%%%%%%%%%%%%

\end{document}